\documentclass{aa}
\usepackage{txfonts}
\usepackage{graphicx}

\def\msol{M$_{\odot}$ }
\def\kms{$\rm km\, s^{-1}$}
\def\cm3{$\rm cm^{-3}$}
\def\Ts{$\rm T_{*}$}
\def\Vs{$\rm V_{s}$}
\def\n0{$\rm n_{0}$}
\def\B0{$\rm B_{0}$}

\def\erg{$\rm erg\, cm^{-2}\, s^{-1}$}
\def\mum{$\mu$m}

\def\Ha{H$\alpha$~}
\def\Hb{H$\beta$~}
\def\ff{${\it ff}$}
\def\Ly{Ly$\alpha$~}

\begin{document}

\title{The symbiotic star CH Cygni}

\subtitle{I. An analysis of the shocked nebulae at different epochs}

   \author{M. Contini, 
          \inst{1,2}
        \
          R. Angeloni\inst{1,2}
         \and
          P. Rafanelli\inst{1}
          }

   \offprints{M. Contini}

\institute{Dipartimento di Astronomia, University of Padova,
               Vicolo dell'Osservatorio 2, I-35122 Padova, Italy\\
              \email{rodolfo.angeloni@unipd.it, piero.rafanelli@unipd.it}
              \and
              School of Physics and Astronomy, Tel-Aviv University,
              Tel-Aviv 69978, Israel\\
             \email{contini@post.tau.ac.il}}

\date{Received - / Accepted -}
\abstract
{We analyse the line and continuum spectra of the symbiotic system CH Cygni.}
{To show that the colliding-wind model is valid  to explain this symbiotic 
star at different phases.}
{Peculiar observed features such as flickering, radio variation, X-ray emission,
 as well as the distribution of the nebulae and shells throughout the system 
are investigated by modelling the spectra at different epochs. 
The models account consistently for shock and photoionization 
and are constrained by absolute fluxes.}
{We find that the reverse shock between the stars leads to the broad lines 
observed during the active phases, as well as to radio and
hard X-ray emission, while the expanding shock is invoked to explain the 
data during the transition phases.}
{}

   \authorrunning{Contini,  Angeloni,  Rafanelli}
   \titlerunning{Shocked nebulae in CH Cyg}
   \maketitle

\section{Introduction}
The symbiotic star (SS) CH Cygni (CH Cyg) has been observed since the 
late '80s of the XIX century.  It was identified as an anonymous red 
semiregular variable  by Gaposchkin's 1952 analysis of Harvard patrol 
plates, and classified as a M6-M7 spectrum star with a 90-100 day period 
and about one-magnitude amplitude. 
This classification seemed to be suitable until 1963 when a hot blue 
continuum appeared, along with a late-type
spectrum and  emission lines of H, He, [FeII] and [CaII], evidence of a 
composite nature. This drove Deutsch (1964) to recognize in CH Cyg the 
brightest SS at visual wavelengths. Since then, active and quiescent phases 
have been alternating without a predictable trend. 

At a distance of 270$\pm$66 pc (according to the Hipparcos parallax measurements
 - Munari et al. 1997) CH Cyg is one of the most studied,
yet poorly understood SSs. It has been observed in the last thirty years 
both photometrically and spectroscopically from radio to X-ray wavelengths 
(Kenyon 2001).

 The basic configuration of the CH Cyg system is still debated, 
despite the large amount of multifrequency observations carried out since 
the first recorded symbiotic activity. 
In particular, it is discussed whether it is a binary or a triple star, and, 
in the latter case, whether composed of  a red giant/white dwarf pair with 
a G star companion (Hinkle et al. 1993) or a red giant/white dwarf pair with 
an additional red giant companion (Skopal et al. 1996). 
The explanation of the total hot component luminosity in a binary star scenario
 might be a crucial problem: specifically, the suggested binary orbital 
parameters (e.g. Mikolajewski et al. 1992) would imply an accretion rate onto 
the white dwarf not in agreement with that requested by the observed hot 
component luminosity. On the other hand, adopting the binary star scenario  
the orbital solution of the triple model (Hinkle et al. 1993) would require a 
low inclination angle.
This is in  contrast  with the jet  appearing  in the 
plane of the sky, implying an accretion disk and/or  an orbital plane virtually
 edge-on. Furthermore, the low inclination implied by Hinkle et al. (1993) 
would result in  an unrealistically low mass (0.2 \msol) white dwarf. 
Finally, the similarity  of the 756 day period of CH Cyg with non-radial 
pulsations in a few semi-regular variables drove Schmidt et al. (2006) to 
interpret this periodic trend more as a non-radial pulsation of the cool 
component rather than a real orbital period.

Several theoretical  papers have appeared presenting SSs as colliding-wind (CW)
 binary systems  (e.g. Kenny \& Taylor 2005).
The observational evidence that actually both the component stars of a SS 
lose mass through strong stellar winds (Nussbaumer et al. 1995), along with 
the presence in CH Cyg of radio, optical and X-ray jets (Karovska et al. 2007),
 have confirmed the importance of shocks for a real understanding of such 
outstanding system. 

CH Cyg is characterized  by the presence of an eruptive white dwarf, showing 
episodes of accretion fed by the high-density wind  coming from the M star, and
 of wind-wind interaction leading to shock waves. 
The nebulae downstream of the shock fronts  show  the spectroscopic marks of  
a turbulent kinematics, resulting in an inhomogeneous and  highly disrupted 
envelope.

 In this paper we review the recent history of CH Cyg,
in the light of a colliding wind scenario (e.g. Girard \& Willson 1987, 
Kenny \& Taylor 2005)  that  has been used  to explain many SSs 
(e.g. Formiggini, Contini, \& Leibowitz 1995, Contini 1997, 
Contini \& Formiggini 1999) and  recently updated including a detailed 
analysis of dust features (Angeloni et al. 2007a,b,c).  

We interpret the observational data throughout both the active and quiescent 
phases, from the 1978 great outburst up to date,  with the aim of investigating
 the physical  and morphological conditions of the system by modelling the 
line and continuum spectra emitted  from the different nebulae.

 We adopt the binary scenario,  considering the $\sim$ 756 day period as 
due to the giant pulsation mode, and the $\sim$ 15 year period as the actual 
orbital period.

 Following  the method used in previous SS investigations, we analyse 
the continuum spectral energy distribution (SED) in different epochs, instead 
of the light curve through the years.
At each epoch we model the data collected from radio to X-ray constraining 
the results by the analysis of the line spectra. The continuum accounts for 
bremsstrahlung and dust reprocessed radiation  from the shocked nebulae, 
radiation from dust shells, as well as  radiation from the stars.  
 Our method is particularly suitable for disentangling the multiple 
contributions of gas and dust in different conditions at different phases.

We focus on the $nebulae$  within the shocked wind-wind interaction regions 
and on the $shells$ ejected by the giant star, constraining the models by the 
observed $absolute$ fluxes.

A short summary of observational history of CH Cyg is presented in Sect. 2. 
The colliding wind model and  our application to CH Cyg are presented in 
Sect. 3. The next sections are  devoted to the analysis of the individual 
active and quiescent phases. As CH Cyg has not been observed regularly in 
the different wavelength ranges, except of the optical one, the SED profiles 
in some epochs present some data gaps:  for instance, there are no UV and
X-ray 
observations for the 1988-1990 quiescent phase, nor mid-IR data for the 
1998-2000 active phase. This  led us to focus on specific physical processes 
in each phase: the basic models which result by disentangling the different 
contributions to the SEDs and to the line ratios at different days during the  
active phase 1978-87 are presented in Sect. 4; in Sect.5, different types of 
IR obscurations occurred during the 1988-90 quiescent phase are indicated; 
near-UV and IR variability during the 1991-95 active phase are discussed 
in Sect. 6, while Sect. 7 (1996-97 quiescent phase) deals with the complex 
nature of the radio emission. The very expanded nebulae and the X-ray 
emission are  modelled throughout the 1998-2000 latest active phase (Sect.8) 
up to date (Sect.9). Concluding remarks and a summary of the results  are 
presented in Sect. 10.
\begin{figure}
\begin{center}
\includegraphics[width=0.45\textwidth]{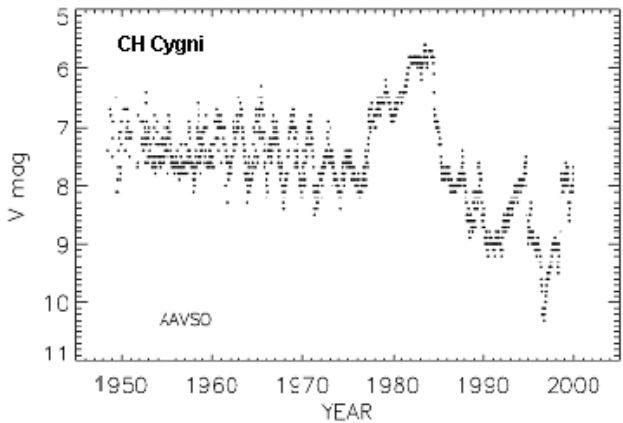}
\caption{AAVSO light curve (10-day means of observations) of CH Cyg from 1948 to 2000.\label{AAVSO}}
\end{center}
\end{figure}

\section{Observational progress}

The symbiotic classification was introduced on the basis of optical
spectral features, however, a real understanding of SSs, and CH Cyg 
in particular, has
been possible only  by space-based astronomy. 

A key-role was played by the long-lived IUE, which opened the UV window
allowing to monitor the temporal evolution of  many emission lines
arising from the nebulae and the continuum coming from the hot star
(see Sect. 4 for a detailed modelling of such temporal evolution).
This great effort  revealed that CH Cyg is  much more
complex than thought before.
 Its puzzling nature triggered a series of new radio observations
(e.g. Taylor et al. 1985, 1986)  attempting to understand  the actual
origin of the nebular emission and to constrain a self-consistent view
in the light of the most accepted scenario at the time.
It was   concluded that the symbiotic behaviour  was due to the dense cool wind
photoionized by the hot star radiation (e.g. Taylor et al. 1988).

The optical and near IR spectrophotometry throughout the years
(e.g. Taranova \& Yudin 1988, 1992; Munari et al. 1996;
Skopal et al. 1996, 1997, 2000, 2002, 2004, 2007) have allowed to follow
 small changes in the emitting properties of CH Cyg,  without
leading to  a substantial breakthrough.

Unexpected results came from the first X-ray observations
(e.g. Leahy \& Taylor 1987) that for the first time made clear that
the energetic of the whole system was hard to explain in term of standard
photoionization models. It became evident that shocks, and in general
hydrodynamic phenomena (e.g. fluid instabilities), play a primary role
in shaping the energy distribution along the whole spectral interval.
Further observational evidence  of stellar wind collision  in CH Cyg
 derived from the most recent X-rays observations 
(Ezuka et al. 1997, Galloway \& Sokoloski 2004, etc).

The IR spectral range is not yet fully explored.
It  was suggested that "symbiotic dust" carries as much information as
symbiotic emission lines (Angeloni et al. 2007c), and the unique
CH Cyg mid-IR spectra recorded by the ISO satellite (Schild et al. 1999)
complicated furthermore  the system overall picture. 
Multiple dust shells, mixed grain chemistry, equilibrium between grain
formation and destruction processes in the turbulent nebulae
(Angeloni et al. 2007a) are worth investigating with the new era of mm-IR
facilities, such as ALMA and Herschel.

 Fig. 1 shows the light curve of CH Cyg throughout many years.
The first ever recorded symbiotic activity  of CH Cyg started in September 1963 and
was over by August 1965, followed by an other period of increased activity
in the years 1967-1970.
Then, CH Cyg experienced a dramatic eruption in 1978 when it became as
bright as to be visible by unaided eye (visual mag. 5.6).
This great outburst ended in 1986 with the emission of bipolar radio and
optical jets (Solf 1987) and by a decline of about 2.5 visual magnitudes
(Taylor et al. 1986). After 1986, the decline in magnitude continued, even
though not regular nor smooth.
Meanwhile, shorter and dimmer outbursts (1991 - 1995, 1998 - 2000) have been
accompanying the fading trend (Fig. \ref{AAVSO}).
Between these periods of activity the spectrum has exhibited
almost exclusively red giant characteristics, and U B V light curves with
significant variations at a level of about 1.5 mag.

Since the beginning of the 2000, CH Cyg persists in a quiescent phase at rather bright magnitude,
despite a 2 mag decline in the U B V light curves during the 2006 June-December
period (Skopal et al. 2007). An apparently anomalous extreme IR fading has
been recently pointed out by Taranova \& Shenavrin (2007).

In the Appendix, we present the references to the observational data we  collected
from the literature. In the quoted works the reader can find the
journal of observations, so in the Tables we have indicated only the spectral
domain and the year, addressing to the specific reference for any further
information or technical detail.

\section{Theoretical framework}
\subsection{CW models for SSs: a short review}

\begin{figure}
\begin{center}
\includegraphics[width=0.45\textwidth]{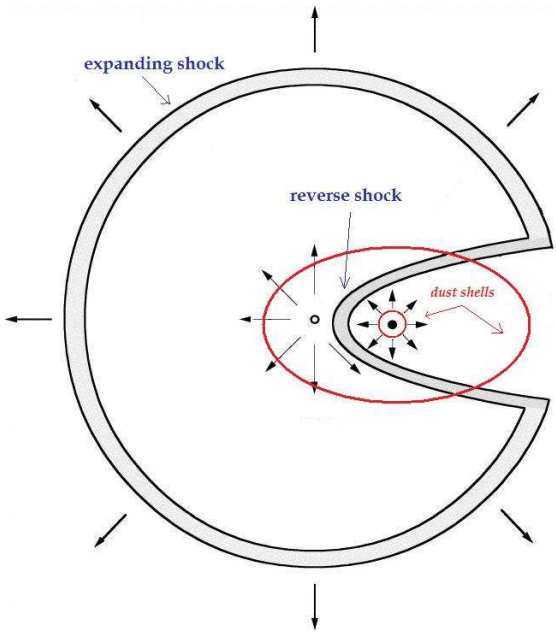}
\end{center}
\caption{A very schematic picture of the CW model and the dust shells
(adapted from Kenny \& Taylor 2005).
\label{schema}
}
\end{figure}

 The first models for the  calculation of radio emission from SSs adopted a single star picture
with a totally ionized wind (e.g. Seaquist \& Gregory 1973)
Then, Willson et al. (1984) reproduced the line profiles  in terms of
the interaction of a low velocity wind from the red giant with a high
velocity wind from a WD in a detached
binary system where both stars have spontaneous winds. The line forming
regions are located  in different portions of the interaction shell and in
the two winds.

A binary  system composed of a cool, mass-losing
star and a hot companion  was also adopted by Taylor \& Seaquist (1984),
reproducing the  radio spectrum
by a simple steady-state model in which a  portion of the stellar
wind is photoionized by the hot companion.

Girard \& Willson (1987)
 presented the colliding wind model which is at  the basis
of further developments. Namely, the winds from  a late giant and from
a hot WD collide between the stars and material is
accumulated along the boundary, forming an initially spherical
shell that  is deformed by the wind of the primary (giant star):
eventually, the shell reaches a steady state configuration. A large amount
of turbulence perpendicular to the mean flow is expected.

Both in the steady state formulation and in the dynamic shell model
the shell is sufficiently thin  to be  considered having zero thickness.

The colliding wind model of Kenny \& Taylor (2005)
 differs from previous ones because assumes
mixing of shocked material from both incoming streams instead than postulating a
self-sustaining contact discontinuity.
The model extends the CW model by the derivation of an adiabatic temperature,
the consideration of radiative cooling, the inclusion of thermal pressures
in the incoming winds, and the treatment of interaction shells of finite
thickness and density. They succeed in reproducing  radio images
and continuum spectra.

Collision of the winds and the  derived
shock network have been  invoked to  resolve  specific discrepancies between observations
and theory during the active phases of a symbiotic phenomenon.
Nevertheless, in order to simultaneously reproduce the UV, optical, and IR line  spectra
as well as the absolute fluxes of the continuum SED from radio to X-ray (Contini 1997, Contini \& Formiggini 2003, Angeloni et al. 2007a,b), the  shocks cannot be neglected  even during the quiescent phases.

\subsection{A schematic model adapted to the detailed calculation of
the  spectra}

 Recent works on a significant sample of dusty SSs have been revealing a 
common trend in the IR spectrum properties, such as similar, recurrent shell 
temperatures and unexpected grain chemistry (Angeloni et al 2007b, 2008 
in preparation).

In order to take into account the importance of dust  in SSs, our model adds to 
the colliding wind binary system 
the dust shells emitted by pulsation of the
cool star.
A very schematic picture of the system is shown
in Fig. \ref{schema}. In this figure the basic shock fronts appear.
The dynamical effects of instabilities, such as fragmentation, will be
explained in the text, as well as evaporation and sputtering of dust grains
in the shells.

 The shocks associated with the colliding winds are  strongest along the
line connecting the two stars.
The interface between the two winds contains two shocks:
one facing the fast
wind, with  \Vs $\sim$ 1000 \kms but low mass flux, and one facing the slow wind,
with \Vs $\sim$  30 \kms  and considerably higher flux corresponding to
mass loss rates of $\sim$ 10$^{-6}$ or 10$^{-5}$ \msol year$^{-1}$.
To a first approximation they can be considered standing shocks.
We will refer to the strong shock facing the WD.
In the expanding region around the hot star away from the direction
towards the cool star, we will refer to the 
expanding shock  outward the system.

 We are interested in the calculations of the line and continuum spectra
throughout the  different epochs,
so we refer
to schematic nebulae: downstream of the  strong standing shock between the star,
downstream of the expanding shock, and downstream of the dust shell edges.
The radiation emitted from the different nebulae and dust shells are
summed up with different weights which depend on the orbital phase,
the pulsation period of the red giant, and on activity of
the WD.

The nebulae downstream of the shock fronts are
compressed by the shock and heated by both the shock and the photoionizing flux from the WD.
Downstream of the reverse shock, the radiation flux from the WD  reaches the shock front
edge of the nebula,
 characterized by very high densities and by a high ionization parameter, being the shock
front close to the WD.
The nebula downstream of the shock expanding outwards the system is characterized
by lower densities. In this configuration, the photoionizing flux from the WD,
with an ionization parameter
 diluted by distance, reaches the edge of the nebula opposite to the shock front.
Furthermore,  the photoionizing flux may be obstructed by some intervening
matter before it reaches the
outer nebula, hence leading to shock-dominated spectra, i.e. showing only the
characteristics of the shocks.

\subsection{The SUMA code}
The line and continuum spectra (free-free and free-bound radiation,
hereafter bremsstrahlung)  emitted by the gas downstream  are calculated by
SUMA (see http://wise-obs.tau.ac.il/$\sim$marcel/suma/index.htm for
a detailed description),
 a code that simulates the physical conditions of an
emitting gaseous cloud
under the coupled effect of photoionisation from an external radiation
source and shocks, and in which both line and continuum emission from gas
is calculated consistently with dust reprocessed radiation
 in a plane-parallel geometry.
In particular, dust grain
heating and sputtering processes are also included, as well as free-free
self-absorption in the radio range, so evident in SSs
(e.g. H1-36, Angeloni et al. 2007b).

Before proceeding  with the modelling of the spectra, we would like to
point out that the plane-parallel condition is an approximation
which is a necessary constraint in  composite models (shock+photoionization).
However, the plane-parallel condition is roughly valid in the interbinary
region where the collision of the
winds is head-on. It is  valid in the circumbinary region
where the radius of the expanding shock created by the head-on-back interaction of
the outflowing winds from the stars  is large enough.
We have  adopted it also for the dusty shells
because they can approach large radii.

The input parameters are: the  shock velocity \Vs, the   preshock density \n0,
the preshock magnetic field \B0, the colour  temperature of the hot star \Ts,
the ionization parameter $U$, the geometrical thickness of the emitting nebula $D$,
the dust-to-gas ratio $d/g$, and the  abundances of He, C,N, O, Ne, Mg, Si, S, A, Fe relative to H.
Solar abundances are adopted (Allen 1973). The distribution of the grain radius downstream
is determined by sputtering,  beginning with an initial  radius of 0.2 \mum. \\

The calculations start at the shock front where the gas is compressed
and thermalized adiabatically, reaching the maximum
temperature
in the immediate post-shock region.
The downstream region is cut in many plane-parallel slabs (up to 300)
with different geometrical widths calculated automatically  in order
to account for the temperature gradient.
 In each slab compression is
calculated by  the Rankine-Hugoniot equations  for the
conservation of mass, momentum and energy throughout the shock front.
In the case where shock and
photoionization act on opposite sides
the geometrical width of the nebula $D$ is an input parameter,
the diffuse radiation bridges the two sides.
 In this case, a few iterations are necessary to consistently obtain the
physical conditions downstream.

The ionizing radiation from an external source is characterized by its
spectrum, which is calculated at 440 energies, from a few eV to KeV.
 Due to the radiative transfer, the
radiation spectrum changes throughout the downstream slabs, each of them
contributing to the optical depth. The calculations assume a steady
state  downstream. In addition to the radiation from the primary
source, the effect of the diffuse radiation created by the gas emission
(line and continuum) is also taken into account
 using about 240 energies to calculate the spectrum.

For each slab of gas, the ionic fractional abundances of each
chemical element are obtained by solving the ionization equations.
These equations account for the ionization mechanisms
(photoionization by the primary and diffuse radiation, and
collisional ionization) and recombination mechanisms (radiative,
dielectronic recombinations) as well as charge transfer effects.
The ionization equations are coupled to the energy equation
when collision processes dominate, and to the thermal balance if
radiative processes dominate. This latter balances the heating
of the gas due to the primary  and diffuse radiations reaching
the slab,  and the cooling, due to recombinations and collisional
excitation of the ions followed by line emission, dust collisional
ionization, and thermal bremsstrahlung. The coupled equations
are solved for each slab, providing the physical conditions necessary
for calculating the slab optical depth, as well as its  line and
continuum emissions. The slab contributions are integrated
throughout the nebula.

Dust grains are coupled to the gas across the shock front by the magnetic field,
and are heated by radiation
from the WD and collisionally by the gas to a maximum temperature which is a function
of the shock velocity,
 of the chemical composition and radius, up to the evaporation temperature
(T$_{dust} \gtrsim$ 1500 K).
The grains throughout shock fronts and downstream might be destructed by sputtering.
Grain emission and absorption processes, which depend on wavelength and on grain characteristics,
may therefore explain the complex structure of the light curves at different wavelengths.

To better understand the  modelling of the SED, one should recall that the frequency of
the bremsstrahlung
peak depends on the shock velocity, while the radiation from an external
source affects the SED
between 10$^{14}$ and 10$^{15}$ Hz. In a consistent way, the dust reradiation
peak shifts to
higher frequencies the higher the shock velocity, while the dust-to-gas
ratio determines
the ratio between the  dust peak  and the bremsstrahlung at
a certain frequency.
The reradiation of dust within the  SS nebulae is, therefore, observable in the IR only for relatively
high dust-to-gas ratios, being the IR range dominated by radiation from the
red giant (Contini 1997, fig. 4; Contini \& Formiggini 2003, fig. 5).

In the following, we adopt a more physical approach by modelling the SEDs
at different epochs instead of light curves at different wavelengths and
translating magnitudes vs. wavelengths in fluxes (\erg)  vs. frequencies.

\begin{table*}
\centering
\caption{The  models\label{tab:sumamod}}
\begin{tabular}{lllllllll}
\hline
\hline
\     & m1$_{shell}$   & m2$_{shell}$   & m$_{600}^a$  & m$_{1200}^a$    & m1$_{exp}^a$ & m2$_{exp}^a$&m3$_{exp}$  \\
\hline
\  \n0 (\cm3)& 1.5e9   & 2.e9       &  5.e7 & 1.e7 &   1.e5  & 1.e7 &2.e7\\
\ \Vs (\kms) & 40      & 60        &  600 &1200  &  150 & 100 & 70  \\
\  \B0 (gauss) & 2.e-3 & 3.e-3     &  3.e-3 & 3.e-3 &  1.e-3 &1.e-3& 1.e-3 \\
\  \Ts (K)     & 1.5e5 & 1.5e5     & 1.5e5&1.5e5 & 3.5e4& 3.5e4&3.5e4   \\
\ U           & 0.5    & 1        & 6 & 2  &  0.01 &0.001 &0.001 \\
\  $D^b$ (cm)    &4.0e14 &3.6e13  & 1.e14 &  1.e14  & 5.e15 & 3.e17& 3.e17 \\
\  $d/g^c$   &  4.e-4  & 4.e-4  & 4.e-4 & 4.e-4 &  4.e-7    &  4.e-7 &4.e-7  \\  \hline
\end{tabular}

\noindent
$^a$ radiation dominated (U$\neq$0);
$^b$ upper limit in radiation-bound models; $^c$ by mass
\end{table*}

\section{1978-1986 active phase: the models}

Since the emission lines better constrain the physical properties of a nebula than the continuum,
we start by modelling the composite line spectra.

\subsection{The line spectra}

Permitted and forbidden lines  were observed along the whole outburst episode.
Hack et al. (1988) reported that in quiescence only \Ha and \Hb appeared as
broad, relatively strong emissions, while the other Balmer lines were faint
absorbed and severely blended with the neutral metallic lines.
The FW0M of \Hb varied from about 400 \kms at the beginning of the outburst to 600-700 \kms
from early 1978 to the middle 1981; then it remained at 600 \kms until mid 1984, increasing
to 750 \kms  in July 1984; in November 1984 and  1985 it reached about 1200 \kms.
At the same epoch, high resolution UV spectra showed a strong and broad
(4000 \kms) Ly$\alpha$ emission, never present at previous epochs.
The origin of the broad \Ly is discussed by Contini et al. (2007).

Before 1985 the overall spectra did not show strong emission features except [SII] 4068+,
 [OI] 6300+ in the optical and OI 1304, 1358, SiIII] 1892, CIII] 1909, and MgII 2800
in the UV (Mikolajewska et al. 1988).
Emission lines of HeI 5876 were present during the whole outburst except in the beginning.
The [OIII] 4363 and 5007 lines  appeared only in November 1984, when the outburst was almost over.
At the same epoch the UV spectrum indicated that several high ionization emission lines
(NV, CIV, SiIV, HeII, OIII]) had appeared (Selvelli \& Hack 1985, Mikolajewska et al. 1988).
The decrease in density at the end of the outburst was indicated by the ratio of
FeII/[FeII] and CIII] 1906/1909.

The UV line ratios suggested electron densities of 10$^8$-10$^{10}$ \cm3, characteristic
of  winds close to the stars. On the other hand, the forbidden optical lines
indicated densities  $<$ 10$^6$ \cm3 and  were therefore emitted from the nebulae
downstream  of the expanding shock front.

Since the structure and evolution of the  nebulae are very complicated,
we adopt three schematic
types  (Tables \ref{tab:sumamod}-\ref{tab:lines}): 1) downstream of a  shock with velocities
of 600 - 1200 \kms corresponding to the reverse shock, 2) downstream of a shock expanding
outwards with velocities $\leq$ 150 \kms which can be either radiation dominated,
namely, the nebula is reached by the photoionizing flux from the WD, or shock dominated, i.e.
the conditions in the nebula are those corresponding to a pure shock,
 and 3) downstream of a shock with velocities of $\sim$ 40-60 \kms and a
high \n0 ($\sim$  10$^9$ \cm3) propagating outwards the red giant star, which represents its own
dust shells. As a matter of fact, the shock velocities of the shells
are $\leq$ 30 \kms, but the velocity increases when the shock propagates throughout the circumstellar environment with a negative density gradient.

We can now explain the evolution of the spectra in the 1978-86 active phase.
 A complete grid of models was run and the models which appear in Table \ref{tab:sumamod}
were selected from those which best fit the observed line and continuum spectra.

\begin{figure*}
\includegraphics[width=61mm]{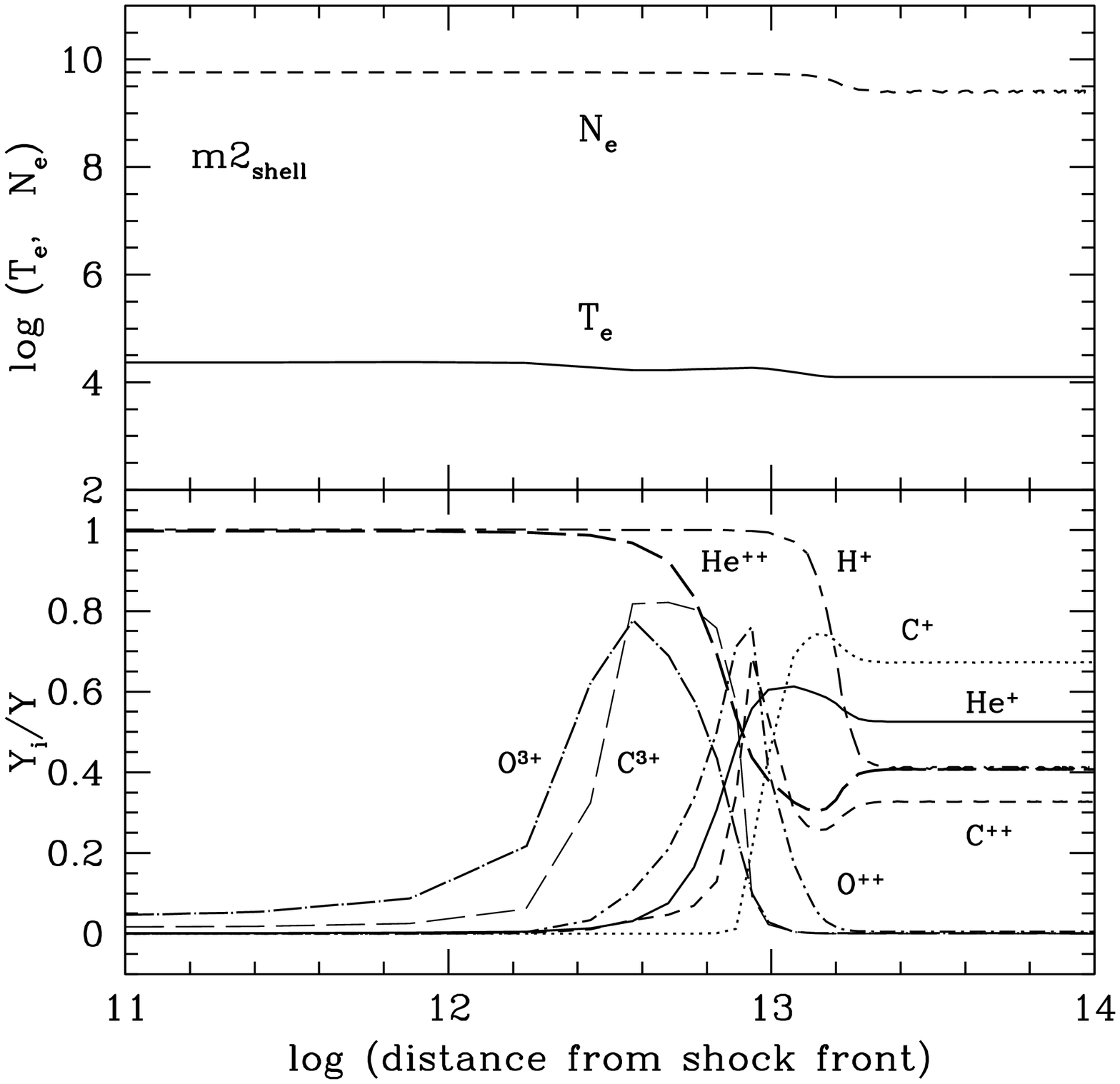}
\includegraphics[width=61mm]{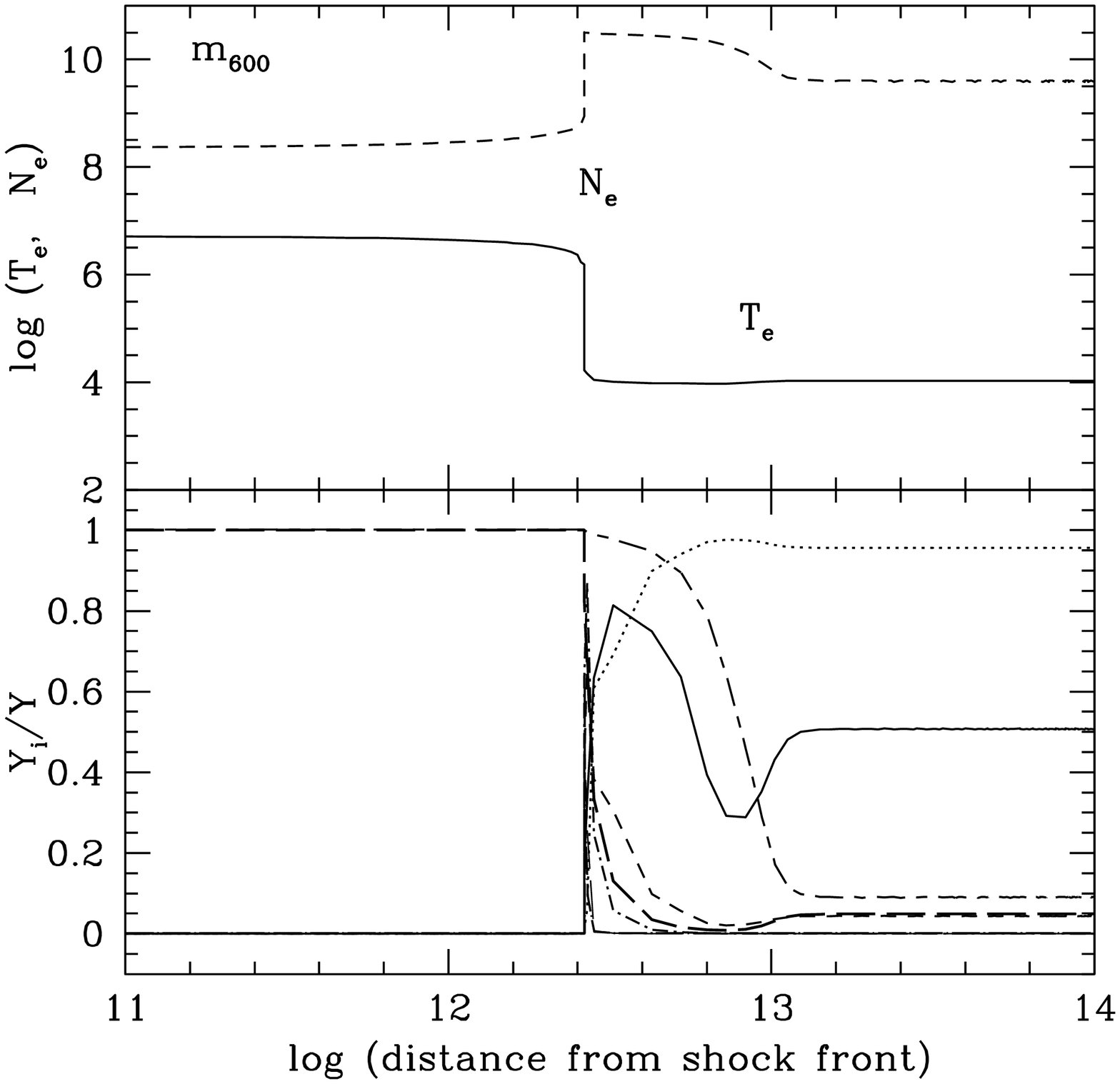}
\includegraphics[width=61mm]{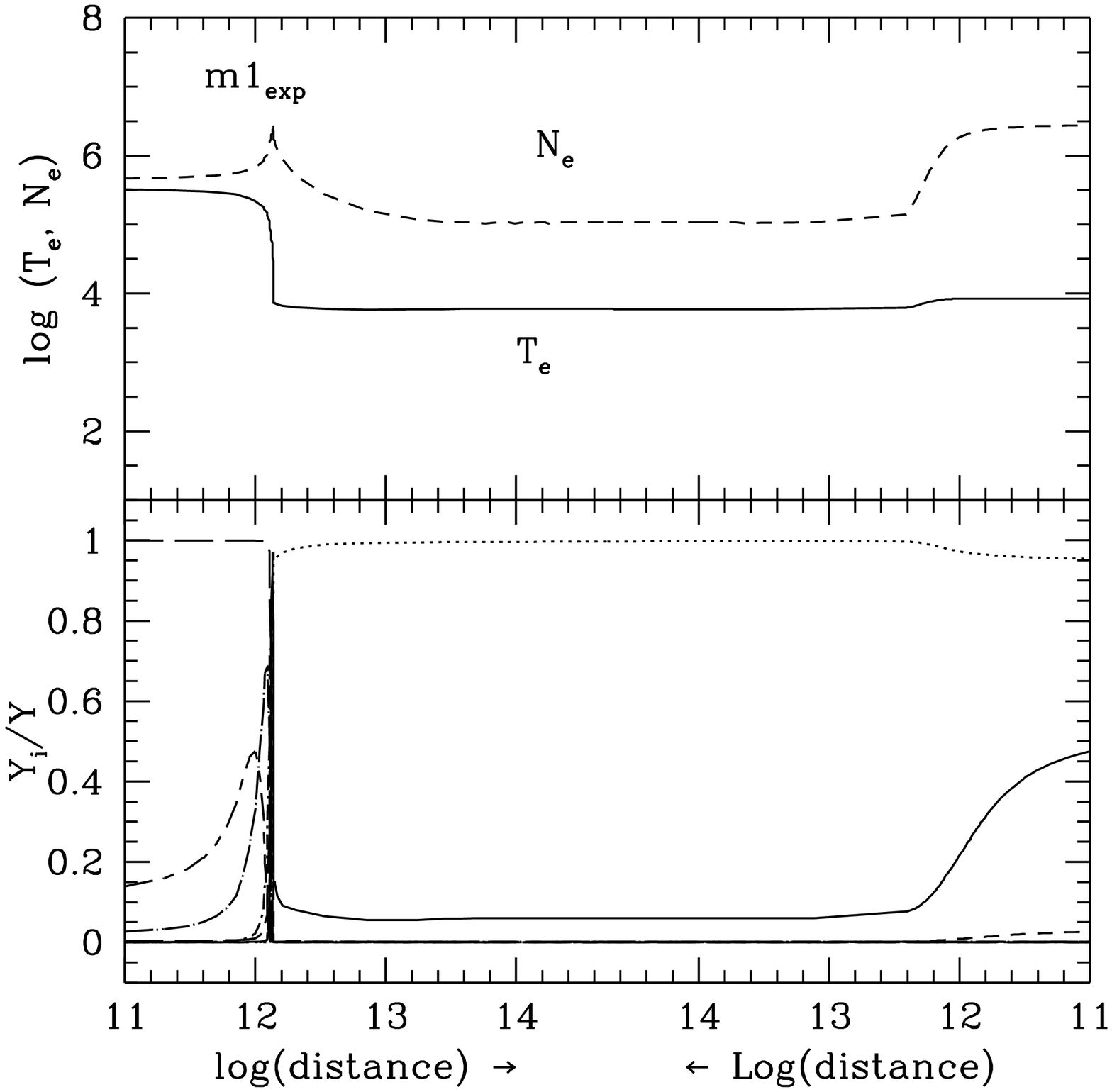}
\caption{Left diagram: The profile of the electron temperature and electron density
throughout the dusty shell  for model m2$_{shell}$  is shown in the top of
the diagram. The fractional abundance of the
most significant ions is shown in the bottom of  the diagram.
Middle: the same as  in the left figure  throughout the nebula downstream of
the reverse shock corresponding tothe model
(m$_{600}$). Symbols are the same as in the  left  diagram.
Right:  the same as in the   diagrams on the left for the expanding nebula 
(m1$_{exp} $). The X axis units are $cm$. \label{fig:ttt}
}
\end{figure*}

In Fig. \ref{fig:ttt} we show the profile of the electron density, electron
 temperature
and of the fractional abundances of the main ions, which explain the spectra.

In particular, each diagram represents the profile of the physical conditions
in the gas downstream of a shock front. We have chosen three very different models.
The diagrams correspond to shocked nebulae. The shock front is on the left.
The diagram on the left of Fig. \ref{fig:ttt} and the middle one show the case in which radiation
from the WD reaches the very shock front edge.
In fact the diagram on the left represents the nebula downstream of a gas+dust
shell ejected by the Mira.
It is characterized by a high preshock density and a low \Vs. The shell propagates towards the WD.

The middle diagram represents the conditions downstream of the reverse shock,
facing the WD. The shock velocity is roughly determined by the FW0M of the line
profile, while the range of the preshock density is dictated by the line
ratios.

The diagram on the right of Fig. \ref{fig:ttt} represents the nebula downstream of the
shock front propagating outwards the system. The X-axis scale is logarithmic
and symmetric with respect to the middle of the nebula  in order to show the conditions
in both  edges with similar high resolution, because, in this case, the shock and
the radiation flux act on opposite edges. The physical conditions on the two sides
of the nebula (the collisional dominated and the radiation dominated)
are bridges by the secondary diffuse radiation from the hot slabs of gas.

The middle diagram of Fig. \ref{fig:ttt} and the left side of the diagram on the right
show the characteristics of a shock dominated regime, namely, the
temperature of the gas decreases from $T=1.5 \, 10^5 (Vs/100 \, km \, s^{-1})^2$
immediately behind the shock-front. However, the recombination coefficients
are low  at high temperatures. Therefore, a long region of high temperature gas
survives downstream, emitting, by the way, X-rays and coronal lines.
The cooling rate is proportional to $n^2$. When the temperature
reaches  about 10$^5$ K the bulk of the UV and optical lines are emitted
leading to a strong decrease of the temperature. The abrupt cooling is
followed by the calculation code by reducing the geometrical thickness of the
slabs (down to even $10^5$ cm) by $(T_{i-1} - T_i)/T_{i-1} <$ 10, where $T_i$ and $T_{i-1}$ represent the temperatures of two
close slabs. After the drop, the temperature  is maintained at about
$10^4$ K by the primary and secondary radiations. Recall that
photoionization alone hardly heats the gas to temperatures $>$ 2-3 $10^4$ K.
In a pure shock case (U=0) the temperatures decreases rapidly  below
 1000 K and the neutral lines are thus strong.

The situation is different in the shell (left diagram) with a relatively
low temperature even close to the shock front. The radiative ionization
rates prevail on the collisional ones which depend strongly on the
temperature, so the picture is very similar to that of a pure photoionization regime.
The density is very high and the optical thickness of the slabs is high enough
to reduce the primary and secondary fluxes at a certain distance from
the shock front. The gas therefore recombines. The case presented for the shell 
is matter-bound due to fragmentation at the shock front.

\subsubsection{Spectra from the reverse shock}

As mentioned before, in the epoch corresponding to the outburst the FWHM of the Balmer lines
evolve
from 400 to 1200 \kms. These lines are emitted from the  nebula  downstream of the reverse shock
bounding the wind-wind collision region between the stars (Contini 1997, Angeloni et al. 2007a).
Model m$_{600}$ is adopted to represent the high velocity shock:
it shows relatively high \Hb, HeII and HeI line ratios (Table \ref{tab:lines}).

The FWHM of the broad line profiles suggests the range of the shock velocity
 \Vs.
High  velocities are characteristic of the
reverse shock between the stars,   a high \n0 is
 dictated by the range of densities in  the  star atmosphere.

Recall that the temperature in the immediate post-shock region is $\propto$ $V_s^2$,
so the temperatures are high  ($\geq$ 2.4 10$^6$ K) for \Vs $\geq$ 400 \kms,
recombination coefficients are  low, and even
compression (n/\n0) downstream is low close to the shock front,  because the 
densities would correspond
only to $\sim$ 4 \n0 (\textit{the adiabatic jump}). The  radiation flux related to the
H  and He lines  and the  continuum   can
be as strong as to be  observable only  at  temperatures $\leq$ 10$^6$ K. 
The densities play a dominant role in the cooling rate
($\propto$ n$^2$, where n is the density)  in order to recover those  temperatures
downstream of a strong shock.
In fact,  the density  must be high enough (\n0 $\geq$ 10$^7$ \cm3)
to reduce the temperature to T$\leq$ 10$^6$ K within the nebula downstream of the
reverse shock.  This nebula therefore should
have a geometrical thickness $D$   smaller than the binary separation (10$^{13}$ - 10$^{14}$ cm).

This shows that \Vs, \n0, and $D$ are the main input parameters  constraining
the choice of the model.
Moreover, a strong tranversal component of the magnetic
field  can reduce compression and a high  dust-to-gas ratios  may speed up
the cooling rate of the gas  through the collision heating of dust grains
(Contini 2004). The ionization parameter and the star temperature affect
particularly the
ionization structure of the gas.

The radiation flux emitted from the gas downstream is
calculated in each slab
 downstream and   integrated throughout
the nebula. The input parameters are  even more constrained by  comparing
the calculated flux F$_c$
with the  flux F$_o$ observed at Earth throughout the continuum SED (Fig. \ref{fig:78sed}).
Recall that    F$_c$ r$^2$=F$_o$ d$^2$, where r is the distance of
the nebula from the  system center and d is the distance to Earth.
For instance,  F$_c$  = 2 10$^8$ \erg  at $\nu$=5 10$^{14}$ Hz is calculated downstream 
of a shock front with \Vs=600 \kms and \n0 = 5 10$^7$ \cm3. For r$\sim$ 2 10$^{13}$ 
$cm$ and
 adopting d= 270 pc,   F$_o$ results $\sim$ 10$^{-7}$ \erg,
in agreement with the observations (Fig. \ref{fig:78sed}).
 In this case the  emission lines  corresponding to  the heavy elements are very weak
because the low temperature ($\leq$ 10$^6$ K) zone in the nebula  is narrow, while
 the Balmer and permitted He lines are high enough to be  observed,
 due to a high  ionization parameter $U$  ($>$ 1).

The parameters are  refined  in order to obtain the best fit
to observations in different epochs,  considering  that \Vs \, increases with time
during the 1978-1986 active phase   and therefore
\n0 must decrease.

\subsubsection{Spectra from the shells}

At the earliest epochs (1978-1984) the  models are constrained by the relative
strength of OI 1304,
1358,  SiIII] 1892, CIII] 1909, and MgII 2800.
The spectra before November 1984  showed relatively  strong low ionization lines which are
explained by  model m1$_{shell}$ (Tables \ref{tab:sumamod} and \ref{tab:lines})
with \Vs=40 \kms. The CIV/CIII] line ratio  depends on the geometrical thickness $D$
 of the nebula namely, the larger $D$, the lower the CIV/CIII] ratio.
Therefore, the observed  CIV/CIII] $<$ 1 suggests   to  represent the shells  by matter-bound
models relative to  a quite extended nebula ($D$=4 10$^{14}$ cm).

 The parameters  describing the shells are chosen considering that dust shell characteristic velocities
are $\sim$ 30-50 \kms and that the densities  must be suitable to the giant atmosphere
($\geq$ 10$^9$ \cm3).
They are slightly  modified  in order to obtain line ratios fitting the
observations. Notice that adopting  low \Vs \, (30-60 \kms) and high densities,
the  temperatures downstream are $<$ 6 10$^4$ K and
the permitted UV lines emitted from the heavy elements are high enough to be observed.

\subsubsection{Spectra from the expanding shock}

The eclipse (Skopal et al. 1996) at the end of 1984 indicated that the system had rotated
and the spectra were hence emitted from a different nebula,
namely, the nebula downstream of the expanding shock.

In November 1984, [OIII] and [NIII] in the optical domain, and several high ionization
lines (e.g. CIV, SiIV) appeared.
The models best fitting both line  and  continuum  spectra are shown in
Table \ref{tab:JD90} where the line ratios are compared with the observations of
Mikolajewska et al. (1988).

 The parameters  corresponding to the expanding shock are  dictated by the FWHM of the
forbidden lines,
by the densities that must avoid collisional deexcitation of the forbidden lines
(e.g. [OIII] 5007),
and by the  consistent fit of the observed continuum SED and of the line ratios.

In particular, the  UV line ratios to \Ly=10  calculated by model m1$_{exp}$ are
compared with the observations  by Mikolajewska et al. (1988, Table \ref{tab:JD90})
at JD 2446086 (20/1/1985).
This model is shock dominated, as the spectrum is emitted from gas in the external side of the
expanding nebula, facing the observer.
Notice that a perfect fitting of the observations is not sensible due to the
approximation of the models.
Therefore we have focused on the line ratios which inequivocally describe  a physical
situation.

The model is constrained by the  CIV:CIII]:CII] line ratios.
The shock dominated model m1$_{exp}$ cannot explain the  observed HeII 1640 line.
 which is better approximated by model m2$_{shell}$.
Also in the optical range model m2$_{shell}$ shows  HeII 4686 and HeI 5876 lines
 as intense as to be observed  (Table \ref{tab:lines}), while
 OI 1304 and SiII 1814  are underestimated.
In conclusion,  also the shells
contributed to the line spectra as well as to the continuum (see Sect. 4.2).

\begin{table}
\caption{Calculated line ratios to \Hb\label{tab:lines}}
\begin{tabular}{ll llll l }
 \hline \hline
\ line & m1$_{shell}$&m2$_{shell}$ &m$_{600}$ &m1$_{exp}^a$ \\
\hline
\ NV  1240   &0.003&0.23 & 0.001 &4.15\\
\ OI 1304  & 0.04&0.02&0.006& 0.006\\
\ CII 1335+  &0.8&1.65&0.15&3. \\
\ OI  1356   &0.42    &0.19&0.06&0.05\\
\ OIV]+      &        &    & & \\
\ SiIV1400 &0.34&1.8&0.02&8.   \\
\ NIV] 1486  &0.03  &0.27   & 0.002&3.  \\
\ SiII 1531  &0.06 &0.03&0.013&0.02\\
\ CIV 1550   &1.0 &7.2  & 0.09&14.     \\
\ HeII 1640  &3.8 &4.7&1.9 &0.06            \\
\ OIII] 1663 &0.25&0.9 &0.05&2.         \\
\ NIII] 1744+&0.1  &0.4 &0.002&1.3        \\
\ SiII 1814 &0.37&0.2 &0.1&0.08           \\
\ SiIII] 1892&0.7 &1.  &0.4&2.5                 \\
\ CIII] 1907+&1.2 &2.  &0.08&5.6              \\
\ CII] 2326   &1.5 &1.2&0.25&3.2          \\
\ MgII 2798  &1.1&0.5 &3.&0.5            \\
\ [NeIII] 3869+&0.008&0.004    &0.001 &0.3     \\
\ [OIII] 4363  &0.008&0.004&-&0.45\\
\ HeI 4471   &0.04&0.02&0.03&0.63         \\
\ HeII 4686  &0.55 &0.65&0.28 &0.005            \\
\ [OIII] 5007+&0.001&0.002& - & 1.47\\
\ HeI 5876   &0.14&0.08&0.1&1.12            \\
\ [OI] 6300+ &0.002&0.0004&0.001 &0.25           \\
\ \Hb $^b$  &7.6e8&1.9e8  &5.8e8 & 0.75        \\
\hline
\end{tabular}

$^a$ shock dominated\\
$^b$ in \erg

\end{table}

\begin{table}
\caption{1978-86: modelling the line ratios at JD 2446086 (20/1/1985) \label{tab:JD90}}
\begin{tabular}{ll ll l  l}
 \hline \hline
\ line & obs$_{6086}$  &m2$_{shell}$ &m1$_{exp}^a$ \\
\hline
\  HI 1215 &10  &10&10 \\
\ NV  1240   &1.6& 0.02& 0.8 \\
\ OI 1304    &3.3&0.02  & 0.01 \\
\ CII 1335+  &0.87&0.13 & 0.6 \\
\ OIV]+SiIV 1400 &0.83&0.15   &1.3\\
\ NIV] 1486  &0.19&0.02&0.43\\
\ SiII 1531  &0.77& 0.002  & 0.02 \\
\ CIV 1550   &1.8&0.58 & 2.7 \\
\ HeII 1640  &1.4&0.4  & 0.01\\
\ OIII] 1663 &0.69&0.007 & 0.4 \\
\ NIII] 1744+& 1.1&0.03  & 0.2\\
\ SiII 1814  &1.23&0.02 & 0.02\\
\ SiIII] 1892 &0.89&0.08  & 0.5  \\
\ CIII] 1907+&1.2 &0.16 & 1.1 \\
\hline
\end{tabular}

$^a$ shock dominated

\end{table}

\subsection{The continuum SED}

In Fig. \ref{fig:78sed} we present the modelling of the observations (Table \ref{tab:refa1})
at some significant days during the 1978-86 active phase.
The top left panel of Fig. \ref{fig:78sed} displays the data on a large frequency range
(from radio to X-ray), while in the top right one the optical-UV range is zoomed in.
The two bottom panels focus, respectively, on the IR and optical-UV  flux variations:
in particular, the bottom left panel shows the  time-evolution of the SED in the near-IR,
characteristic of a significantly variable absorption by dust.

Model results are  plotted over the data in the top panels of Fig. \ref{fig:78sed}.
The models  are
cross-checked by  the line spectra previously  analysed (Sect.4.1) and
constrained by the strong self-absorption in the radio range (top left panel).
The high velocity models m$_{600}$ and  m$_{1200}$  are needed to explain the
X-ray datum (top left panel).
The models characterised by \Vs $>$ 600 \kms
 reproduce the  Balmer line FW0M,  but  lead to  UV-optical lines  negligibly weak,
so they  were not included in  Tables \ref{tab:lines} and \ref{tab:JD90}.

Dust is collisionally heated by the gas to high temperatures in a strong shock velocity regime
and sputtering reduces the grain sizes; the smaller the radius, the higher the temperature
reached by the grains.
Reprocessed  radiation by dust downstream of the  shock fronts appear in the top left diagram
of Fig. \ref{fig:78sed}. Model m$_{600}$  can reproduce both the optical and radio continuum slopes.

The dusty shells which  fit the data  in the optical-UV domain at  maximum of activity
(JD 2444937-2445720 = 28/11/1981-20/1/1984) contribute also to the CIV and CIII]  lines in agreement with
observations. Dust in the shells  is mainly heated by radiation from the hot star:
this was already found by previous investigations (Munari et al. 1996, Angeloni et al. 2007b,c).
Throughout expansion, the dust shells gradually  fragmented by interaction with the environment,
 may reach  the external edge of the  binary system ($>$ 10$^{14}$ cm),
becoming circumbinary and playing a primary role in the complex obscuration episodes of CH Cyg.

At JD 2446212 (26/5/1985) and 2446776 (11/12/1986 - Fig. \ref{fig:78sed}, right panels) at the end of the active phase
a {\it flat} component appears in the near-UV SED. This is interpreted by model m1$_{exp}$ which
represents the nebula downstream of the shock propagating outwards the system.
Model m1$_{exp}$  is  shock dominated (U=0)  and  explains most of the line ratios observed by
Mikolajewska et al. (1988) at JD 2446086 (20/1/1985 - Table \ref{tab:JD90}).
The slopes at JD 2446212 (26/5/1985) and 2446776 (11/12/1986)
can  also be fitted by the same m1$_{exp}$ model in the radiation dominated case,
adopting a relatively
low U. However, in this case, the calculated line  ratios  are lower
than  observed.  Recall that the line spectrum and the  continuum were observed at different,
although close, days. This is a further evidence that if  the medium surrounding
the system is clumpy enough, the radiation
flux throughout its path from the WD to the external shock can be flickering.

During the brightness minimum in 1985, CH Cyg showed X-ray emission (Leahy \& Taylor 1987).
Simultaneous radio observations with the Very Large Array (VLA) revealed a radio outburst
and expanding jets (Taylor et al. 1986).
The X-ray datum  in Fig. \ref{fig:78sed} was taken at the end of 1984, therefore constraining the models
at that time.
For  earlier times, when no data are available, Leahy \& Taylor suggested a flux lower
by a factor of $\sim$80. We agree with them that the very large column density of the
downstream nebulae led to large absorption in the 0.05 - 1.5 keV range.

\subsection{Radii of shells and nebulae}

Comparing the bremsstrahlung  calculated at the nebula with the data observed at Earth
we obtain the factors $\eta$ = r$^2$/d$^2$, where r is the distance of the nebula from
the system centre and d is the distance to Earth.

\begin{figure}[ht!]
\begin{center}
\includegraphics[width=0.5\textwidth]{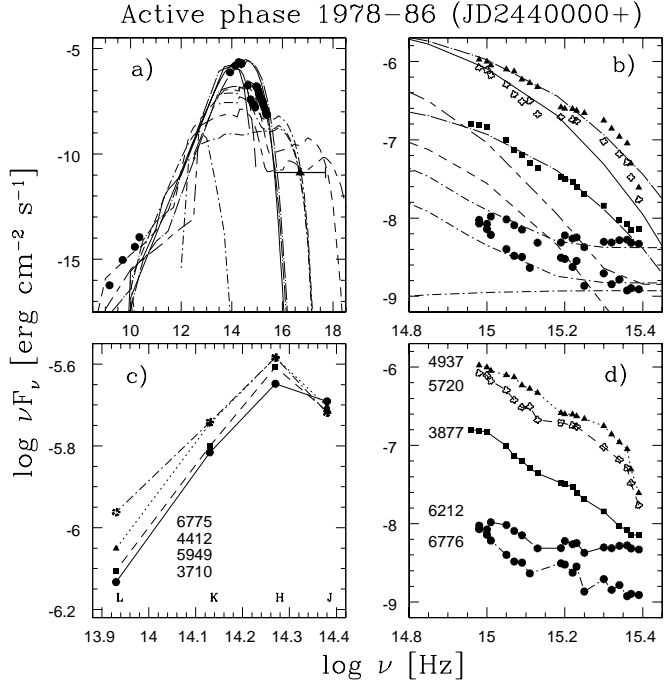}
\caption{Modelling the active phase 1978-86. Top pannels: comparison of the models with
data from radio to X-ray (left) and a zoom in the optic-UV range (right).
Thick-solid line: bb (2000 K); long dash-dotted lines: m2$_{shell}$; solid line: m1$_{shell}$;
short-dashed lines: m$_{600}$; short-long-dashed lines: m$_{1200}$;
short dash-dotted lines: m1$_{exp}$.
Bottom pannels: comparison of time variations in the IR and in the
optic-UV range at different days.\label{fig:78sed}}
\end{center}
\end{figure}

The reverse shock (model m$_{600}$)  was hence located at a  radius of 2 10$^{13}$ cm.
 The top right panel of Fig. \ref{fig:78sed} points out that the bremsstrahlung from the shell in the
optical-UV range, represented by model m2$_{shell}$,  can explain the data at two different days,
JD 2443877 (3/1/1979) and JD 2444937 (28/11/1981). The calculated radii are 2.88 10$^{13}$ and 9 10$^{13}$ cm, respectively,
to which would correspond a shell expansion of  $\sim$ 6 10$^{13}$ cm  in 2.9 years.
This, in turn, would correspond to a velocity $<$ 7 \kms, much lower than the shock
velocity (\Vs=60 \kms),
 unless the two shells were ejected at different times.
The latter hypothesis seems more realistic,
as the pulsation frequency of the cool component is actually $\sim$756d.

The expanding shock reached a  radius of 8 10$^{16}$ cm. The geometrical thickness $\Delta$
of  the cool region (T$\le 10^4$ K) within the downstream nebula is constrained by the
relatively strong  self absorption of free-free radiation in the radio.
The optical thickness  $\tau$  is $\propto$ n$^2$ $\Delta$, where the density n
($\leq$ 10$^{6}$ \cm3
considering compression downstream)  is constrained by the slope of the bremsstrahlung:
a larger n would lead to a  steeper bremsstrahlung in the UV-optical range.
This allows us to give a lower limit  to the geometrical thickness of the cool region :
$\Delta$  $>$ 2 10$^{15}$ cm.

\section{1987-1990 quiescent phase: the obscuration effect}
\begin{figure}
\begin{center}
\includegraphics[width=0.5\textwidth]{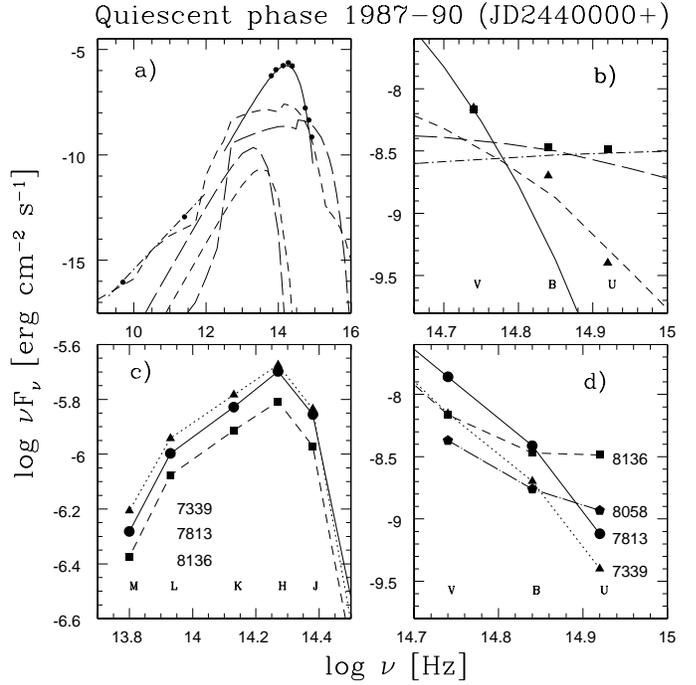}
\caption{Modelling of the quiescent phase 1987-90. 
The figure is structured as Fig. \ref{fig:78sed}. Top panels: Solid line: bb (2000 K); 
short-dashed lines: m3$_{exp}$; long-dashed lines: m2$_{exp}$; dot-dashed lines : model m1$_{exp}$. 
The models refer to Table \ref{tab:sumamod}\label{fig:88sed}.}
\end{center}
\end{figure}

\subsection{The line spectra}
Taranova \& Yudin (1992) stated that the optical spectra  taken in 1988 did not reveal 
any emission 
lines (Mikolajewski et al. 1990) or revealed only very week \Ha and \Hb lines 
(Andrillat 1988, Bopp 1990). In 1989, before the next outburst, an emission line 
spectrum with 
FeII, [FeII], [SII] 4068, and faint one-component \Hb was lying upon numerous absorption lines. 
The spectrum can be roughly explained by the expanding nebula represented by model m1$_{exp}$ 
with  \Vs=150 \kms. At JD 2448122 (18/8/1990), Tomov \& Mikolajewski (1992) observed numerous emission lines from the shell which can be explained by model m2$_{shell}$.

\subsection{The continuum SED}
In Fig. \ref{fig:88sed} we present the SEDs observed during the quiescent period 1987-1990 
between the two active phases 1978-86 and 1991-95. 
Fig. \ref{fig:88sed} is structured as Fig. \ref{fig:78sed}. 
In the top right panel the V data are fitted by the bb flux from the  cool star, 
while the B and U  
data showed a flat slope that can be reproduced by models m1$_{exp}$ and m2$_{exp}$, 
corresponding to different preshock densities (Table \ref{tab:sumamod}). 
So we constrain the models by the line ratios, recalling (Sect. 5.1) that the optical 
spectra observed  
in 1988 did not reveal any emission lines or very week \Ha and \Hb. 
This  excludes the shock dominated model m1$_{exp}$ that was used to fit the flat slopes 
at the 
end of the 1978-86 active phase and the line spectrum at JD 2446086 (20/1/1985).

The models which best fit the slopes at JD 2448136 (1/9/1990) and JD 2447339 (26/6/1988), m2$_{exp}$ and m3$_{exp}$, 
respectively, are both shock dominated with \Vs $\leq$ 100 \kms and relatively high \n0, 
which lead to negligible line emission downstream. 
They indicate that the outflowing shock is propagating through a denser medium, 
perhaps a residue of matter ejected during previous eruptions.

Finally, let us compare the near-IR fluxes during the 1978-86 phase with the ones during 
the 1987-90 
phase (bottom left panels of Figs. \ref{fig:78sed} and \ref{fig:88sed}, respectively): 
while in the former case the slopes showed a dust absorption clearly wavelength-dependent, 
during the quiescent phase minimum the IR flux decreased equally throughout the whole 
near-IR range, 
suggesting an obscuration effect by an external shell.

\section{1991-1995 active phase: the black body radiation from the WD}

A new active phase began in December 1991 with a large flux rise in U, a smaller one in B 
and a negligible one in V. This trend was sustained until March 1992. 
Then, from the beginning of 1993 the U band rose again until summer 1994. 
The eventual drop in summer 1995 informed that the outburst was running out. 

\begin{figure}
\includegraphics[width=44mm]{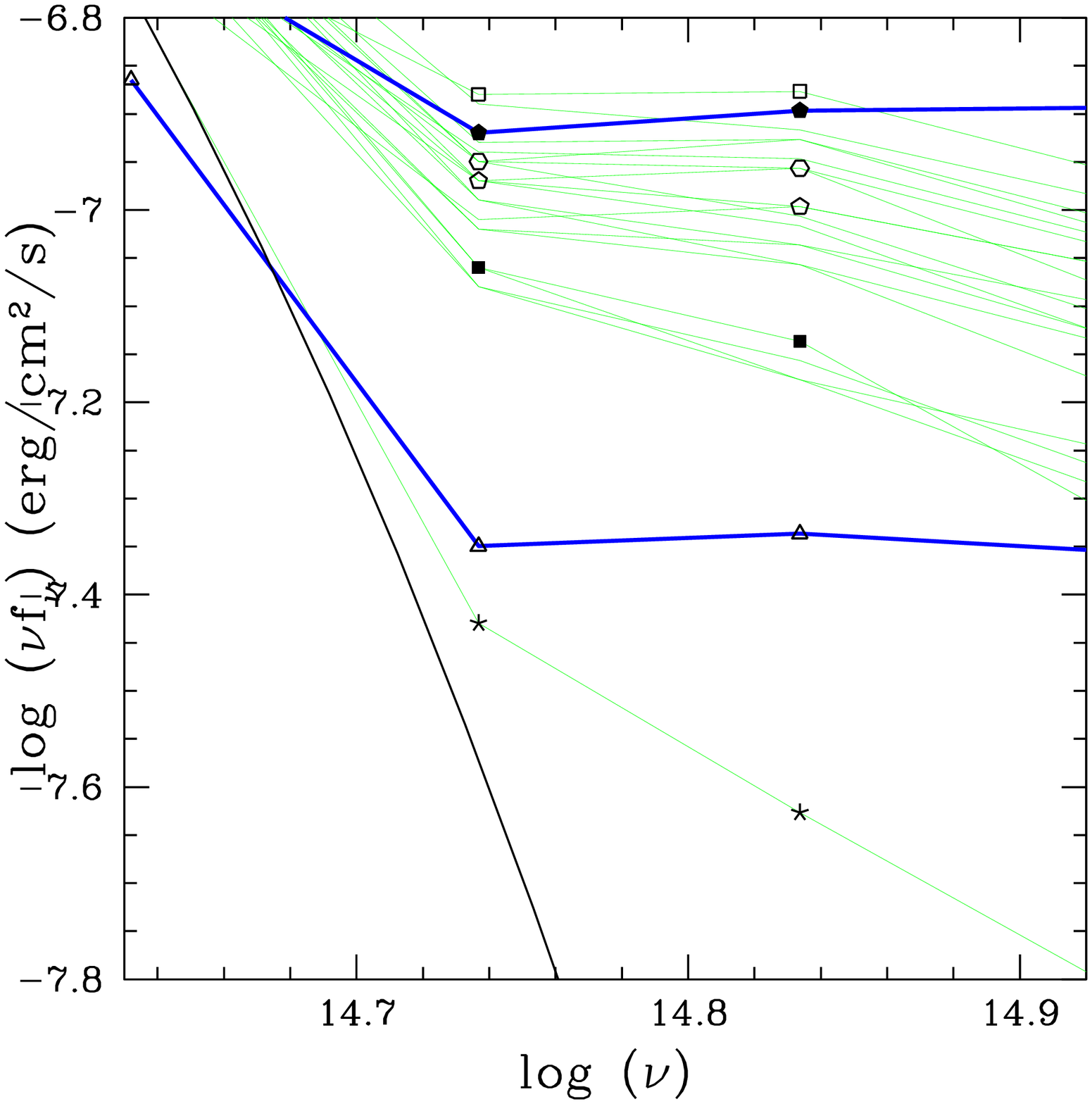}
\includegraphics[width=44mm]{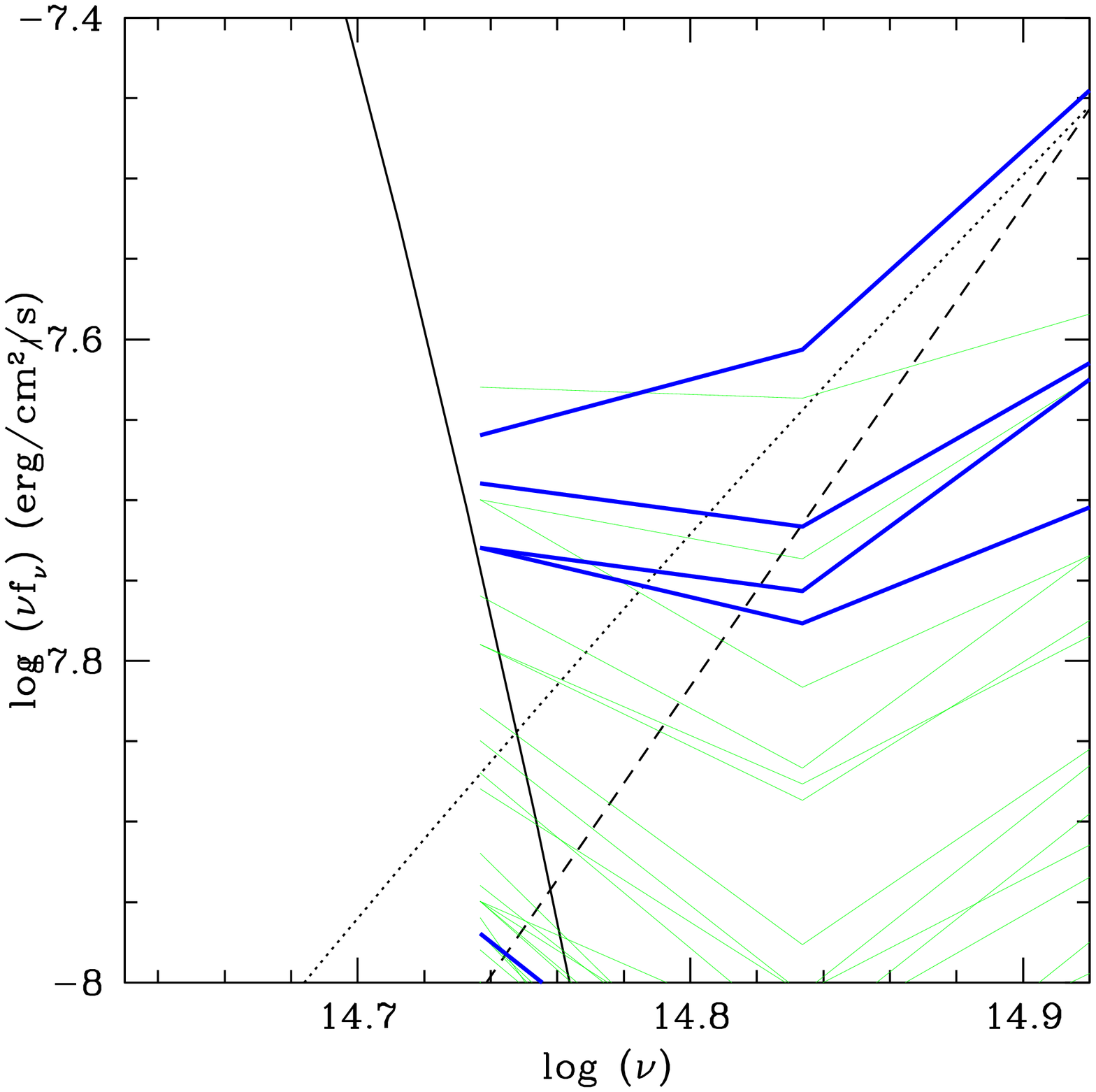}
\includegraphics[width=44mm]{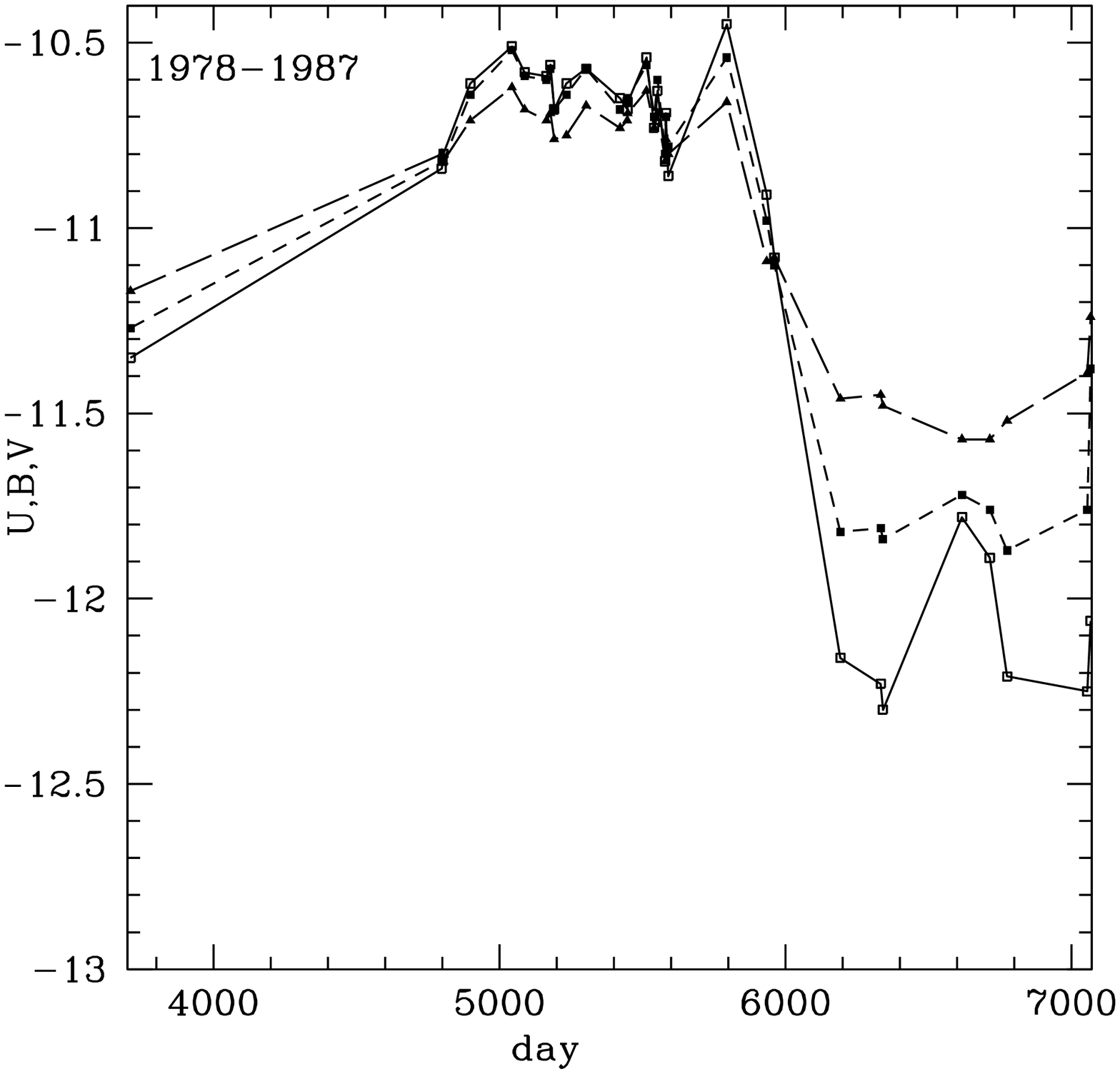}
\includegraphics[width=44mm]{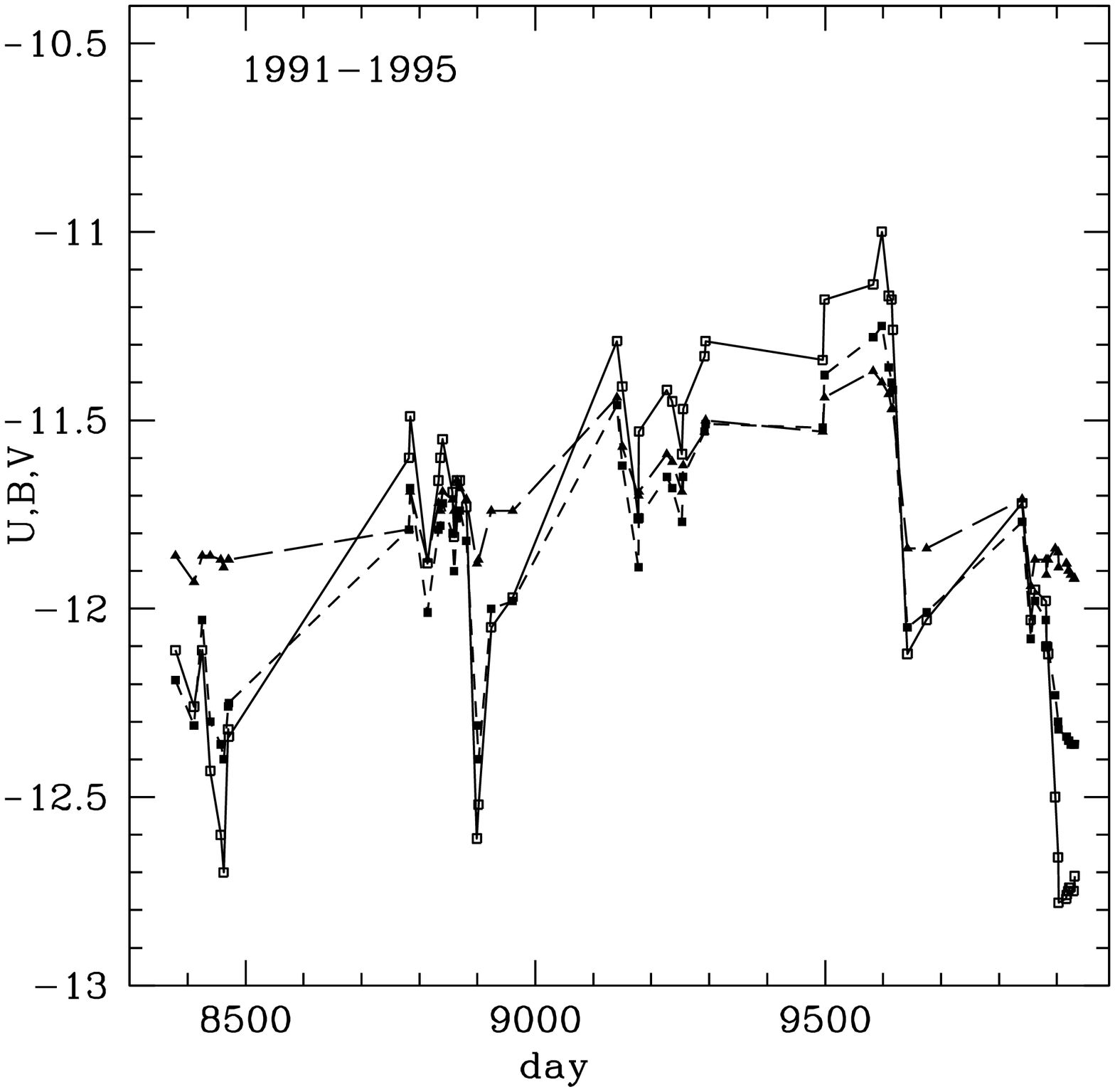}
\caption{Top. Left: the SEDs in different days (1978-87) from Taranova \& Yudin 
(1988). 
Right: the SEDs in different days (1991-95) from Munari et al. (1996). 
Thin lines represent the epochs before the U band maximum, thick  lines refer to 
later epochs. Bottom (JD 2440000+). Left: the light curves  during 1978-87 (in \erg  \AA$^{-1}$). 
U: open squares connected  by solid lines; 
B: black squares connected by short-dashed lines; 
V: black triangles connected by long-dashed lines. Right: the light curves during 1991-95. \label{fig:comparison}}

\end{figure}

It is worth noting that during this 1991-95 high state, the bb continuum flux from the hot star 
became gradually detectable (right panel of Fig. \ref{fig:comparison}). As a matter of fact, the appearance of a 
hot continuum source with T$_{eff}$=150,000 K was taken as the indication 
of the beginning of the outburst by Skopal et al. (1996).                                                                                                            
To obtain an initial   hint about the system  during this active phase,
 we  compare  the evolution of the  slopes in the U-B-V  during the
1978-87 and 1991-95 active phases in Fig. \ref{fig:comparison}.
 An impressive difference can be noticed comparing the right and left panels: in particular, 
the left one  shows that in the first active phase the trend in the 
UB range cannot be explained by the bb flux from the hot star which vice-versa is evident 
(right diagram) during  the  1991-95 outburst. 
The direct contribution of the hot star to the U band is shown at  the 1991-95 active phase 
maximum (JD 2449598 $\equiv$ 2/9/1994).  In Fig. \ref{fig:comparison}, the low frequency tails of the black body curves correspond 
to \Ts =100,000 K (dashed line) and \Ts=35,000 K (dotted line). These temperatures represent 
the maximum and minimum one, 
 respectively, during and after the outburst (Corradi et al. 2001). 
The slopes of  both bb  fluxes in the  UB  frequency domain are slightly 
different than observed,
 and are even less fitting the data before (thin lines) and after maximum (thick lines). 
We suggest that the WD  was hidden from a direct view in  the 1978-86 active phase, 
while it became visible during  the 1991-95 maximum.                                                                                                            

The two deep minima in U recorded at the end of September 1992 and in October 1994 
lasted 81 and 127 days, 
respectively. They were attributed to the eclipse of the active component by the cool one in the 
light of the triple star model (Skopal et al 1996), but they are hardly explained by the binary 
interpretation which ascribes the 756d periodicity to a non-radial pulsation of the M giant. 
Fig. \ref{fig:comparison} (bottom panel) shows that the eclipses of September 1992 just before JD 2448900 (4/10/1992) and 
of October 1994 
involved only the B and U bands, the WD and the shocked nebula being eclipsed  most probably by   
dust shells. The V band in fact  was not involved as  it recovered  its range of magnitudes 
that were shifted during the bursts by the  contribution of the nebulae. 
Similarly, the relatively weak maxima immediately after the "eclipses"  were due to emission 
from the shocked nebulae. 
We suggest that  the minima  were due to absorption by the dusty shells that are ejected in tune with 
the pulsation of the red giant. Recall that the dusty shells are highly fragmented therefore 
this effect cannot be  observed periodically.

\subsection{The line spectrum}
At JD 2448094 (21/7/1990), a dramatic change led to broad hydrogen absorption profiles and at JD 2448095 (22/7/1990)
CH Cyg displayed faint emission in \Ha with corresponding velocities around -67 \kms. 
Then, in August 1991, the appearance of absorption terminal radial velocities 
of -1000 and -900 \kms for \Ha and \Hb, respectively, might have been a consequence 
of the rotation of the system. 
This would have allowed us to see the reverse shock in front of the WD. 
Eventually, the absorptions turned into emissions with velocities up to 1800 for \Ha 
and 1600 \kms for \Hb: these velocities disappeared at the beginning of the 1992. 
From JD 2448883 (17/9/1992) the \Hb  emission increased, whereas no absorption lines were observed. \\
High velocities had been observed also during the previous active phase (Sect. 4) and 
there explained by the reverse shock model.

The spectral region around  [OIII] 5007  showed complicated structures consistent 
with more components. During the outburst (1992 onwards)  strong emission 
lines of FeII, [FeII], [SII], and H were observed, while [NeIII]3869+ and [FeIII] 4659-4670  
were still faint, increasing later. Also, [OIII]4363 and [OIII] 5007+  were relatively 
faint and varied rapidly during the active phase. 
The overall scenario, with multiple emission line components and structured line profiles, 
may be qualitatively understood by Richtmyer-Meschkov (R-M) and Kelvin-Helmoltz (K-H) 
instabilities
(Graham \& Zhang 2000). R-M is an interfacial instability between two fluids of 
different density driven by shock waves. 
The  K-H instability leads to the consequent fragmentation of matter at the shock front 
(Contini \& Formiggini 2001). 
The clumpiness of the nebula  might then contribute to flickering episodes.

On September 1992, the M-giant  was even  more apparent due to the eclipse. 
The red part of the spectrum  was dominated by the M-giant molecular  spectrum, 
overlayed by strong \Ha, [OI]6300, and  HeI 5876 emissions. 
There  were simultaneous emission lines of [SII], [NeIII], and [FeIII]; 
notice that the HeII 4686 and 1640 lines were not present during the outburst. 
We suggest that the line spectrum is most probably  emitted from a disrupted shock 
dominated expanding nebula (m1$_{exp}$) characterised by low HeII lines (Table \ref{tab:lines}).

In May 1993 the UV emission  line spectrum showed  \Ly, OIV], SiIV 1400, CIV 1550, OIII] 1666, 
SiIII] 1892, CIII] 1909, CII] 2325, [OII] 2470. 
These lines are reproduced by the  model for the 
expanding nebulae m1$_{exp}$. 
Moreover, numerous FeII multiplets appeared, together  
with MgII 2796, 2803 doublet,  and sharp emission lines, ArIII 2187, 2191. 
In August 1994, a sharp absorption cutting the emission line profiles at \Vs=150 \kms \, 
most likely shows 
that  the expanding nebula is facing the observer.
In fact, the strongest emission came from the line core while a very broad and smooth absorption  
up to -1600 \kms  was seen.

\subsection{The continuum}

\begin{figure*}
\centerline{\Large Active phase 1991-1995 (JD 2440000+)}
\includegraphics[width=60mm]{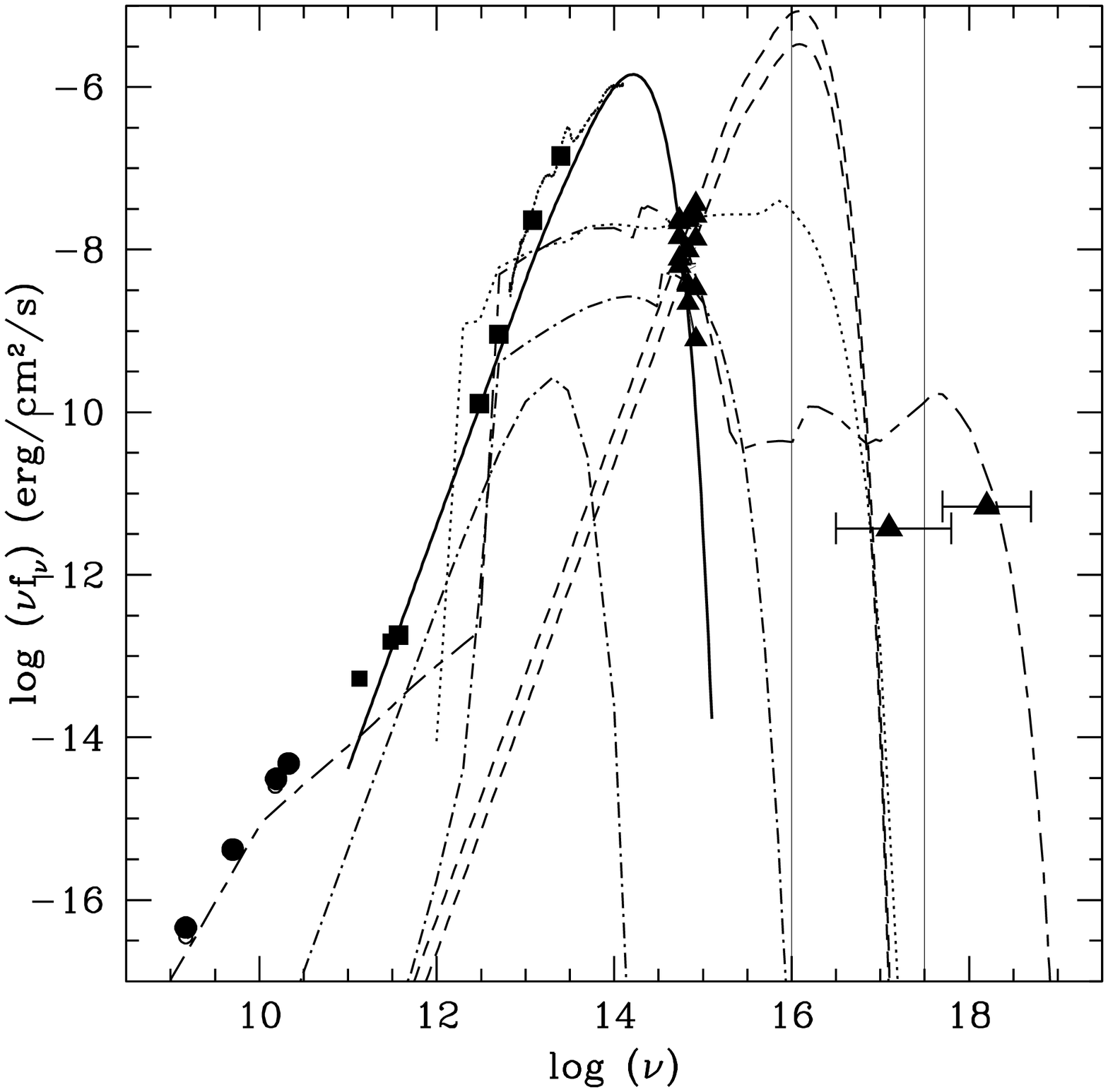}
\includegraphics[width=60mm]{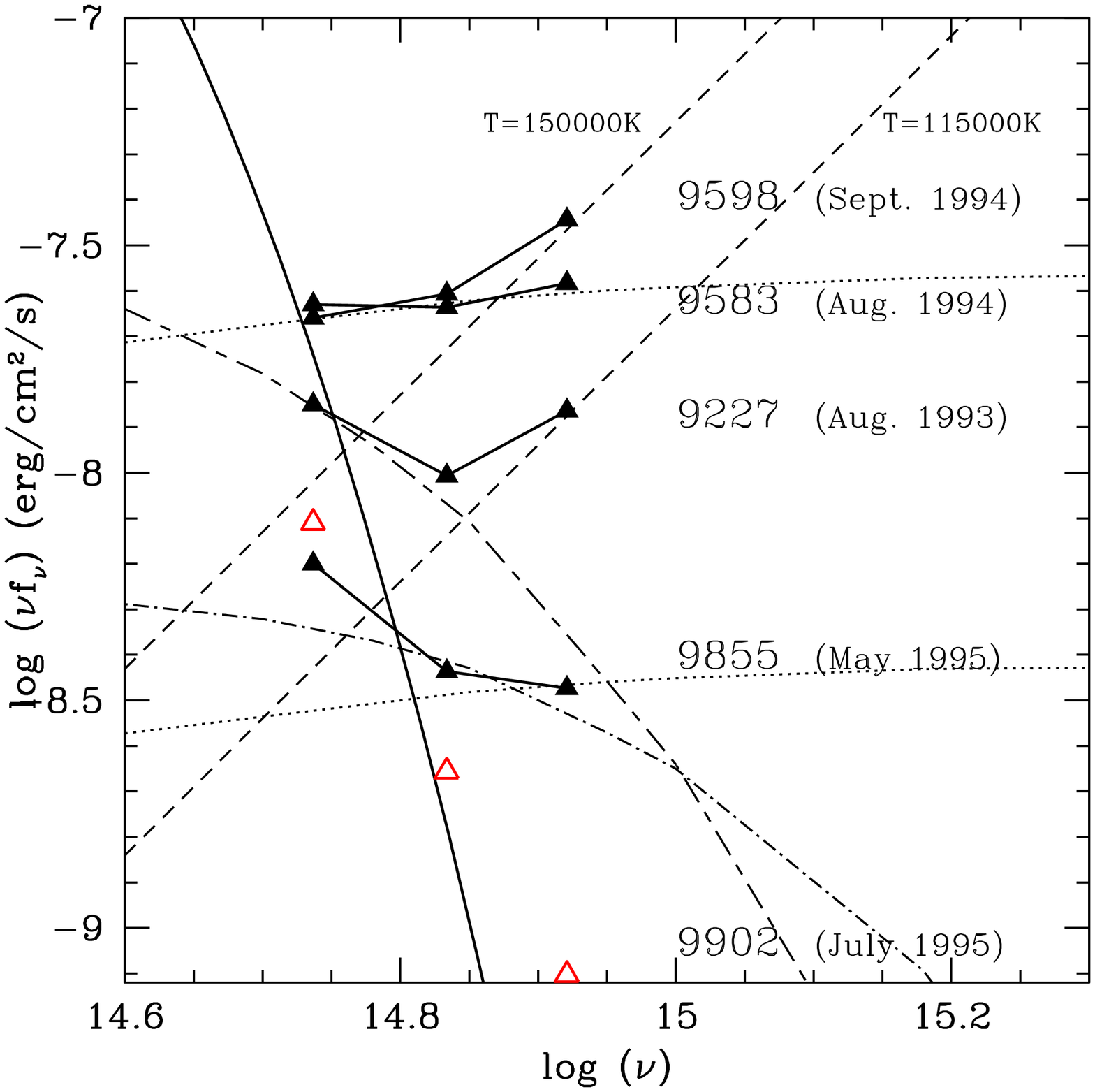}
\includegraphics[width=60mm]{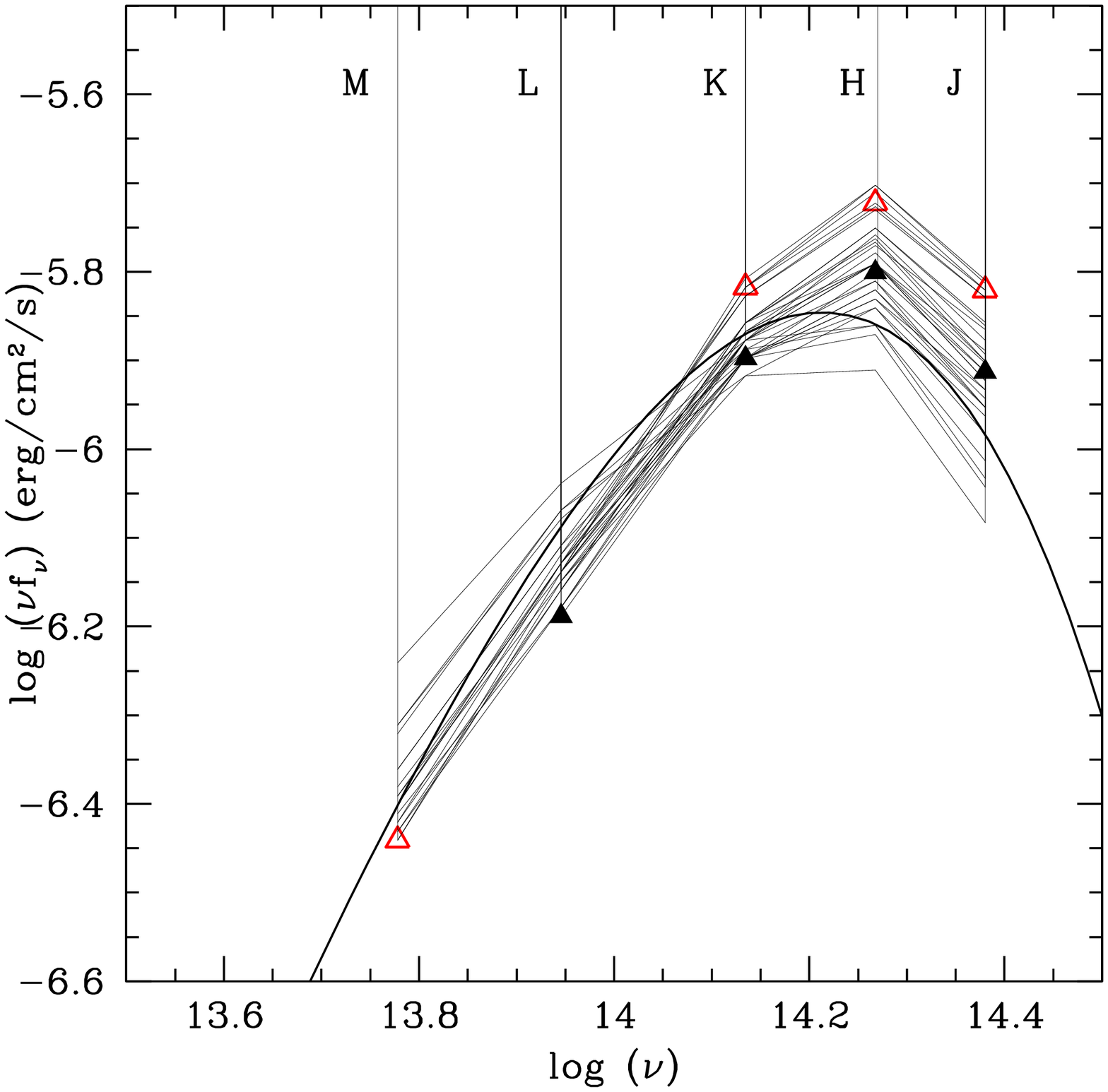}
\caption{Modelling the  1991-1995 active phase on a large frequency range
(left diagram).
Dotted line: model m1$_{exp}$;
dot-dashed: m2$_{exp}$; long-short-dashed: m$_{1200}$; thick solid: bb (2000 K);
short-dashed: bb from the WD (see text); thick dotted line: data from ISO;
 black circles: JD244922, black squares: JD 2448818
(Skopal et al. 1996), small  black triangles : data from Munari et al (1996);
large black triangle: X-ray.
Thin vertical lines indicate the frequency region of high X-ray absorption.
Middle diagram : optic-UV range.  Black triangles :   JD are
labelled for each epoch; empty triangles : JD 2449902.
Diagram on the right: variability in the IR. Data from Munari et al. (1996).
 Black symbols show the slopes of the lines;
empty triangles : JD 2449902.
\label{fig:91}}
\end{figure*}

The SEDs of the continuum  are shown in Fig. \ref{fig:91} diagrams. 
The models which  appear in the left diagrams, covering the whole spectral range, were selected by the consistent fit of 
the data in the optical-UV range (zoom in the middle diagram), of the radio and of the X-ray data.

We start by presenting the modelling in the optical-UV range because it is more constraining.
Around  maximum at JD 2449598 (2/9/1994), the  flux SED  corresponding to the sum of the expanding 
shock (m1$_{exp}$) bremsstrahlung and the bb  flux from the hot star (\Ts =150,000 K)
reproduces  the slope of the U-B-V SED. At JD 2449583 (18/8/1994) the bb component was not visible. 
This may be explained by the fragmented nature of the shells which screen the flux from 
the WD, confirming Eyres et al. (2000) claim that flickering cannot be observed 
during the eclipse.

Similarly, the  flux SED  which best reproduces the trend of the data at JD 2449227 (27/8/1993) results 
from the sum of model m$_{1200}$ with the WD bb flux: recall that broad lines were still 
emitted in 1993. In May 1995 we observed only the expanding shock corresponding to 
model m2$_{exp}$.

Fig. \ref{fig:91}, middle diagram shows that the bb flux from the WD  grew by a factor of 2.75 
between JD 2449227 (27/8/1993) and JD 2449598 (2/9/1994), corresponding to an increase of the WD  temperature 
from $\sim$ 115,000 K to 150,000 K.

We can now examine the data on a large scale, from radio to X-ray. 
The radio slope between 1992 and 1993 
as observed by MERLIN, VLA, and JCMT (Skopal et al 1996) is once again explained by bremsstrahlung 
with a strong self-absorption at longer wavelengths. 
The best fit is obtained with a model corresponding to \Vs=1200 \kms in the reverse shock. 
This model also explains the data in the optical-UV range and the hard X-ray. 
Models m1$_{exp}$ and m2$_{exp}$ give a negligible contribution in the radio. 
Skopal et al. (1996) proposed a  velocity of $\sim$ 310 \kms in order to match the 
contemporary angular radius observed by MERLIN:  this discrepancy is easily explained 
as we deal with  disrupted nebulae downstream  of head-on shocks  in a highly turbulent 
regime.

The soft X-ray datum by ASCA was modelled by Ezuka et al. (1998) as two optically thin 
thermal plasma emissions. 
We find (Fig. \ref{fig:91}, left diagram) that the soft X-ray is emitted downstream of the expanding 
shock, while  the hard X-ray would correspond to the reverse shock between the stars 
with a velocity of 1200 \kms.

An impressive systematic evolution of  the JKLM band flux appears in the right
 diagram of Fig. \ref{fig:91}, indicating that dust is involved in the obscuration process.\\

Finally, by exploiting the $\eta$ factors defined in Sect. 4.3 we can calculate for the reverse shock  
a radius of 4 10$^{13}$ cm  and for the expanding nebulae which best fit the data a corresponding  radius of 10$^{16}$-3 10$^{17}$ cm. Similar dimensions of the system were found by HST imaging in the 1998-2001 active phase by Eyres et al (2002) after about 5 years. Adopting en expanding velocity $\leq$ 100 \kms the radius increment is relatively small ($<$ 2 10$^{15}$ cm).

\section{1996-1997 transition phase: the contribution of radio synchrotron radiation}

\begin{figure}
\includegraphics[width=0.5\textwidth]{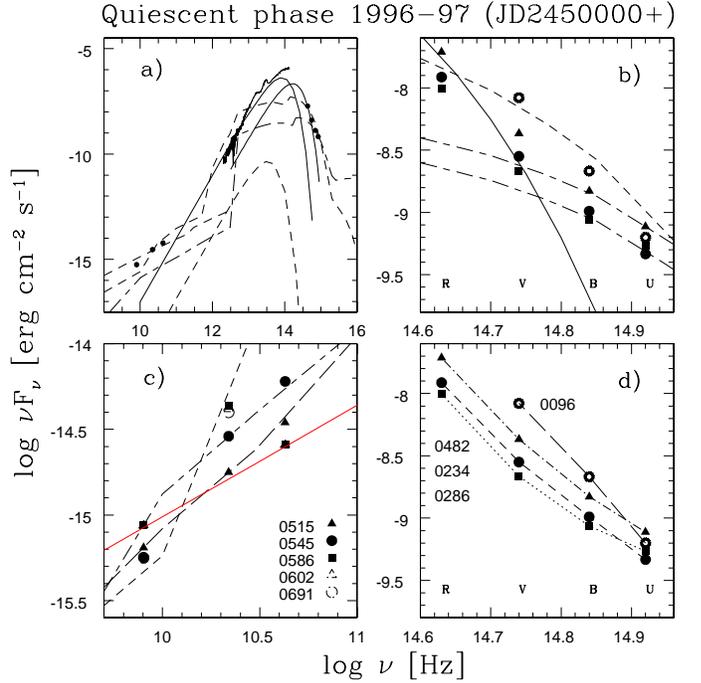}
\caption{Modelling the SED of the transition phase 1996-1997 on a large frequency range (left top) and
in the optic-UV range (right top). The figure is structured as Fig. \ref{fig:78sed}.
 Long-short dashed lines:
m$_{1200}$; dashed lines: m3$_{exp}$; thick solid line: bb flux (2000 K); thin solid line:
dust bb radiation at 1000 K. Diagram on the  bottom left:
variability of the radio flux in 1997.  The data by Karovska et al (1996)  are
connected by thin solid lines; thick solid (red) line: synchrotron radiation corresponding
to $\alpha$=-0.35; short-dashed lines:
m3$_{exp}$; short-long-dashed lines: m$_{1200}$; long-dashed lines: m$_{600}$. \label{fig:96}
}
\end{figure}

During the quiescent phase 1996-97 some interesting radio observations were taken between 
March and October 1997 by Karovska et al. (1998). The data (Fig. \ref{fig:96}, top left), along with an 
intriguing flux variability, followed the trend of an absorbed bremsstrahlung as was found 
in previous phases adopting a high velocity model (\Vs=1200 \kms). 
A particular investigation of the radio range is shown in Fig. \ref{fig:96},  bottom left diagram. 
The slope of the data at 8 and 43 GHz observed between May and August 1997 
could be fitted by synchrotron emission created at the shock front by the Fermi mechanism in the 
non-relativistic case (Bell 1978) with a characteristic spectral index $\alpha$ = -0.35. 
This is not surprising because non-thermal emission was also invoked by Crocker et al. (2001) 
in order to explain the radio data between 1985 and 1999, 
adopting a magnetic field  of the order of 1-3 mG.
Interestingly, we have reached the same values of the magnetic field  (\B0 = 1-3 mG, 
Table \ref{tab:sumamod}) by consistent modelling of line and continuum spectra
in the different phases.
Nonetheless, the datum at 22 GHz is highly underpredicted by the synchrotron radiation 
and may refer to a completely different 
emission process, e.g. bremsstrahlung from the downstream nebula. 
A high velocity model is in fact requested by the optical-UV data, as presented in Fig. \ref{fig:96} 
(top diagrams).

\begin{figure}
\begin{center}
\includegraphics[width=0.45\textwidth]{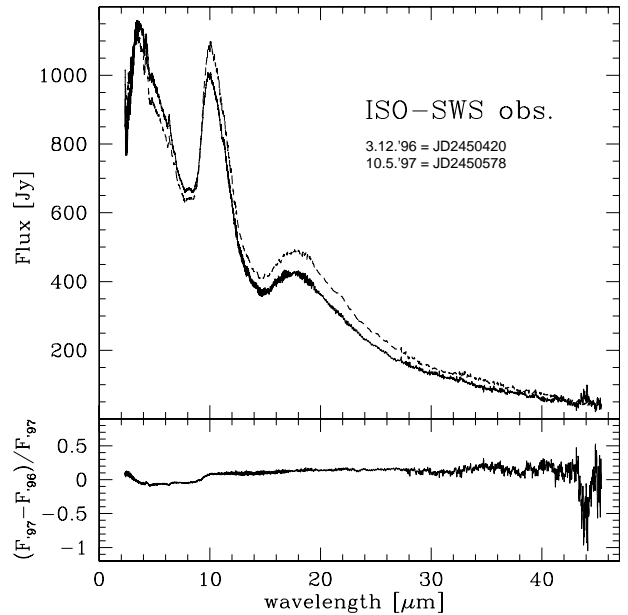}
\caption{The two ISO-SWS spectra. Top panel: the spectra displayed in arbitrary units. 
It is noticeable the variations in the 9.7 \mum\ silicate profile. 
Bottom panel: the relative ratio of the fluxes, showing that the main variations seems to 
be due to the silicate emissions, and not to the underlying dust continuum\label{fig:ISO}.}
\end{center}
\end{figure}

Regarding the infrared range, it is worth reminding some results of 1997 ISO observations, 
unfortunately still no fully exploited. 
Schild et al. (1999), who did a preliminary analysis of the spectra, found no strong 
emission lines but [OI]63 \mum, and strong OH and weak H$_2$O molecular emission 
on top of a rapidly decreasing continuum. This continuum can be fitted by a single 
dust shell model ($T_d \sim$ 1000K, Fig. \ref{fig:96}), in contrast with other SSs which seems to 
display multiple dust shells with characteristic radius and temperatures 
(Angeloni et al. 2007c).\\
The dust chemistry was quite surprising, since weak PAHs (at 6.3 \mum) coexisted  
with dominant amorphous silicate (at 9.7 and 18 \mum). 
Unfortunately, a comparative study of dust spectroscopy capable of investigating 
the grain formation and destruction processes throughout both the active and 
quiescent phases is still not possible for lacking of data 
(see Angeloni et al. 2007a for a comparison of emission lines and silicate dust features, 
from ISO spectra, in a various sample of SSs including CH Cyg). 
ISO-SWS observed the system twice during this quiescent phase, and in Fig. \ref{fig:ISO} 
we show a comparison of the two recorded spectra. 
As one can notice, the maximum flux variation seems to be in correspondence of the strong 
silicate emissions, while the underlying dust continuum appears not so sensitive to the 
environment variation due to the giant pulsations. 
How this can be related to the actual dust formation and processing is far  from understood. 

\section{1998-2001 active phase: jets and X-rays}
\subsection{The line spectra}
As noticed by Eyres et al. (2002) in the HST imaging, the very inner regions of CH Cyg 
contributed mainly to continuum, and very little to the line emission. 
In our interpretation this is not surprising since the nebulae emitting the line spectra 
would correspond to the expanding nebulae that are moving towards the outside of the system.

Balmer lines \Ha, \Hb, H$\gamma$,  all double peaked,  lay  upon a broad feature 
1200 \kms wide  confirming that  a fast reverse shock of \Vs =1200 \kms was 
present within  a  small radius  $<$  10$^{13}$ cm. 
The fast shock can explain also the X-ray  Chandra observations of 2001 
at  $\sim$ 0.26 and 2 keV  reported by Galloway \& Sokolowski (2004). 
Indeed, the temperature in the immediate post shock region corresponding to 
that \Vs, is 2.16 10$^7$ K, decreasing downstream to $\sim$ 1.5 10$^7$. 
The cooling rate is not so strong because recombination coefficients are low at such 
high temperatures.  

The HST images by Eyres et al. (2002) revealed strong [OIII], which can be explained 
by the expanding nebulae m3$_{exp}$  modelled by fitting the SEDs.

In September 1998 the OI, HeI, HeII, FeII, and TiII  were probably emitted 
from  the shells. [OIII] 5007  was also evident, while   [OIII] 4363 absent both 
in and out the eclipse. We have run some further  models corresponding to expanding 
nebulae (ME$_{exp}$ and ME1$_{exp}$) which, in agreement with the observations,  
predict relatively low [OIII] 4363 (Table \ref{tab:jet}).  The input parameters are
shown in the bottom of Table \ref{tab:jet}. 
The models are shock dominated (U=0).
The dust-to-gas ratios adopted in these models are 4 10$^{-4}$ by mass.

Corradi  et al. (2001) indicated strong [OII] and [NII] lines in the 1999 images. 
These strong  low 
ionization lines are characteristic of  shock dominated  models, e.g. 
ME$_{exp}$ and ME1$_{exp}$, which might occur from the  collision of  
the expanding jets  with the surrounding
ISM.  The radius of these shocked nebulae are calculated by fitting the continuum SED. 
The $\eta$ factors  lead to a  radius $ >$ 3 10$^{16}$ cm. 

\subsection{The continuum SED}
\begin{figure}
\centerline{\Large Active phase 1998-2000 (JD 2450000+)}
\includegraphics[width=44mm]{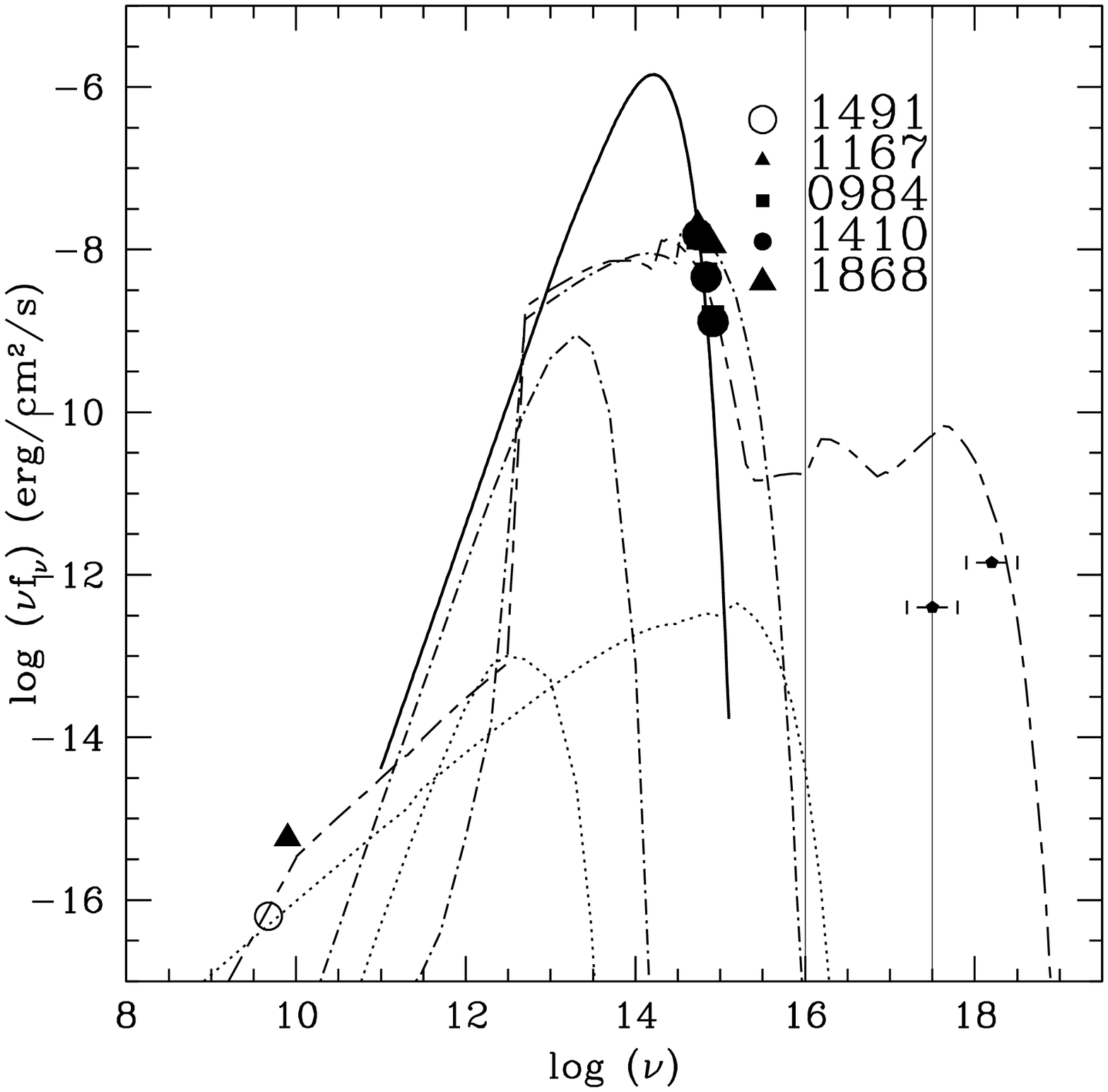}
\includegraphics[width=44mm]{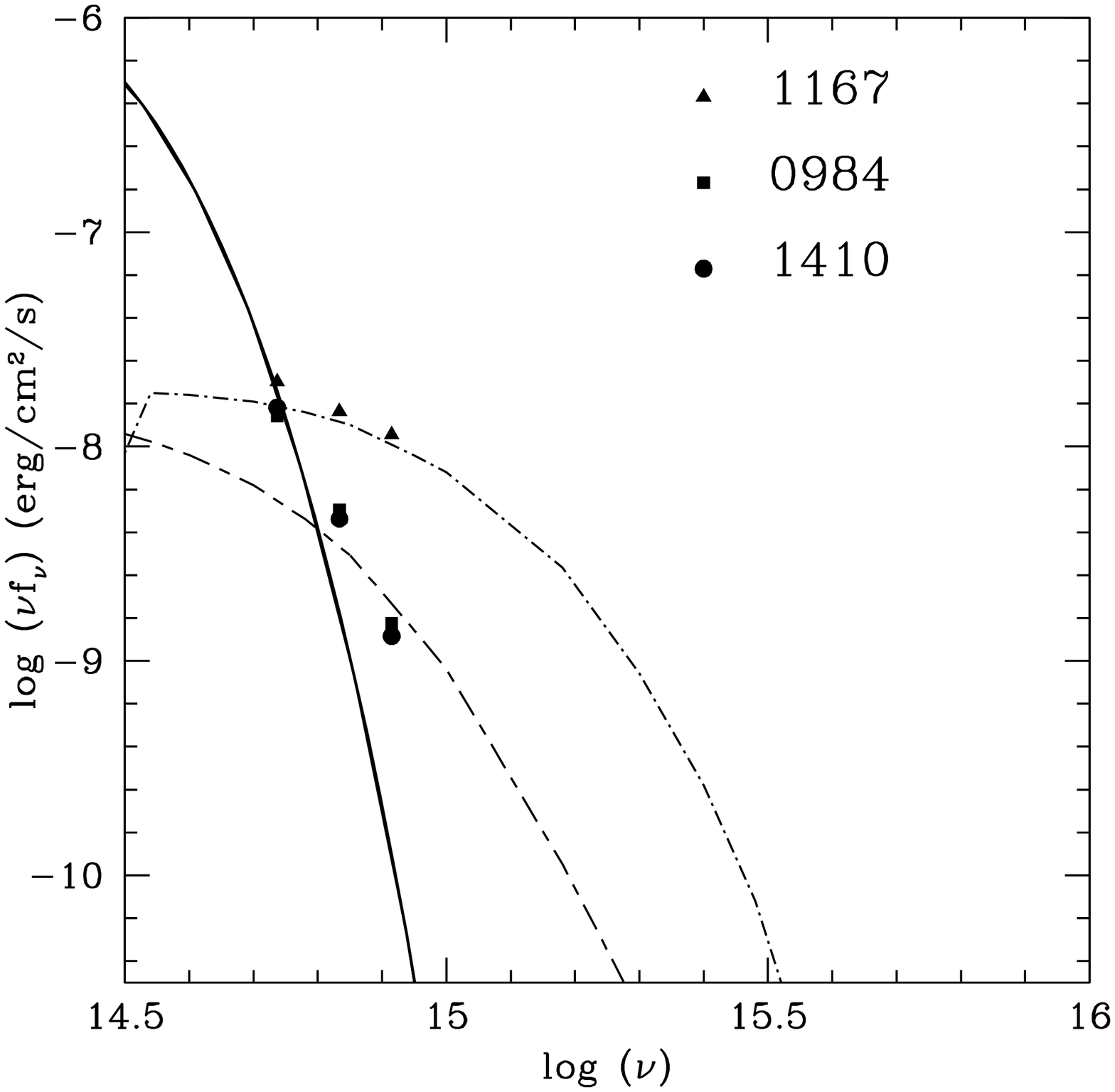}
\caption{The  1998-2001 active phase. Left diagram: modelling the SED  from radio to X-ray. 
Right diagram: the optical frequency range. Long-short-dashed lines: m$_{1200}$; 
dash-dotted lines: m2$_{exp}$; dotted lines: ME$_{exp}$; solid line: bb (2000 K). 
Large black triangle:  data from Karovska et al (2007). 
The vertical thin lines show the range of strong absorption in the X-ray domain.\label{fig:2000}}
\end{figure}

In Fig. \ref{fig:2000} we present the modelling of the 1998-2001 active phase. 
Some significant UBV data are taken from Eyres et al (2002, Fig. 2). Fig. \ref{fig:2000}  (left diagram) 
shows that the best fit of the UBV slopes are obtained by  model m2$_{exp}$ in December 
1998 during the maximum, while at later epochs
the slopes are reproduced by model m$_{1200}$. 
This can be explained considering that at maximum 
the relative  position of the different system components (stars, nebulae, shells) is 
such that we 
cannot see the reverse shock, which appears only after some months. 

The  fitting of the models to the data in Fig. \ref{fig:2000} (right diagram) leads to similar factors $\eta$ 
for the two models. However, we know that the radius of the reverse shock is much smaller than 
that of the expanding shock by at least a factor of 100.  
This question is resolved invoking the filling factor as a key parameter. 
The radii of the expanding shell at maximum of the active phase and  of the reverse shock at 
later epochs result of 1.3 10$^{12}$/\ff ~ cm and of 1.4 10$^{12}$/\ff ~cm, respectively, 
leading to corresponding filling factors  of $\sim$ 0.001 and 0.1, to recover more
sensible radii (cfr. Sect. 4.3). 
Small \ff are suitable to the fragmented  character of  matter   throughout 
the expanding shock deriving from  dynamical instabilities  
e.g. R-M and K-H (Sect. 6.1 and  Contini \& Formiggini 2001).

Fig. \ref{fig:2000} (left diagram) shows also that  the  $d/g$ ratio adopted by model ME$_{exp}$ 
is just a lower limit,  because the flux corresponding to reradiation by dust  
is constrained by the SED,  definitively  dominated by  bb from the cool star in 
that wavelenght range. The low velocities of these models  indicate that the 
expanding nebulae swept up a large amount of interstellar matter, whose $d/g$ 
actually corresponds to $d/g$  $\geq$ 4 10$^{-4}$. This matter probably existed 
before the start of this outburst, and might be considered as a sort of remnants 
from the former bursts which have then propagated throughout the ISM.

\begin{table}
\caption{Line ratios to \Hb in the extended nebulae (jets)\label{tab:jet} }
\begin{tabular}{ll ll l}
 \hline \hline
\ line & ME$_{exp}$ &  ME1$_{exp}$\\
\hline
\ [OII] 3727+& 40.& 19.\\
\ [OIII] 4363 & 2.6 &2.18\\
\ HeII 4686   & 0.016 & 0.027\\ 
\ [OIII] 5007+& 35.3& 25.7\\
\ HeI 5876    & 0.4 & 0.58\\
\ [OI]  6300+ &0.37 & 0.40\\
\ [NII] 6584+&8.9 & 6.42\\
\ \Hb (\erg) & 0.0013&0.0025\\
\hline
\ \Vs (\kms) & 70 & 100 \\
\ \n0 (\cm3)& 5.e3 & 5.e3 \\
\ \B0 (gauss)& 1.e-3 & 1.e-3 \\
\ D  (cm)   & 1.e15 & 1.e15\\
\hline
\end{tabular}
\end{table} 

\section{From 2001 up to date}
\begin{figure}
\includegraphics[width=0.5\textwidth]{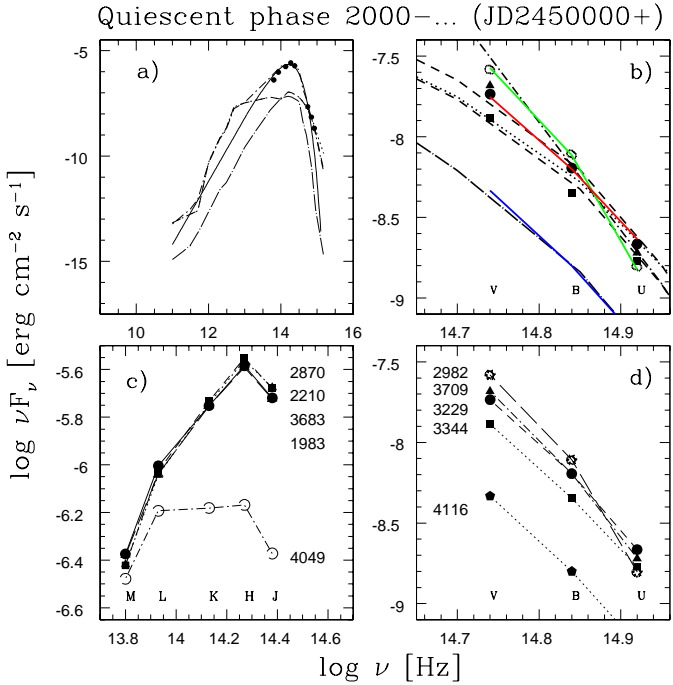}
\caption{
Red line  : the SED  at maximum of  U; green line : the SED at maximum of V;
blue line : the SED towards minimum.
Short dashed line: model m1$_{exp}$; 
short-dash-dotted line: w1 m1$_{exp}$ + w2 bb$_{2000}$, where w1=10$^{-12.5}$ and w2=10$^{-14.4}$ 
are the relative weights, respectively; long-dash-dotted: w1 m1$_{exp}$ + w2 bb$_{2000}$ where 
w1=10$^{-13.35}$ and w2=10$^{-16.0}$ are the relative weights, respectively; 
dotted: w1 m1${exp}$ + w2 m3$_{exp}$ where w1=10$^{-12.5}$ and w2=10$^{-18.9}$ are the relative 
weights, respectively.\label{fig:2007}}
\end{figure}

Since 2001, CH Cyg persists in a quiescent phase recovering from the latest high state, 
at a rather 
bright magnitude (U$\sim$10 or less, V=7-8, B=8.7-9.4). 
During the last months it showed a $\sim$ 2mag 
decline in the optical bands (2006 June-December, Skopal et al. 2007) and a most significant, 
anomalous, IR fading (Taranova \& Shenavrin 2007).

In April 2004, high-resolution optical data by Yoo (2006) 
detected double-peaked emission components 
for the \Ha and \Hb profiles, with an interesting V/R $<$1, where V and R stand for the 
intensities of the blue and red wings, respectively. This has been taken as a proof, along 
with the vanishing of the blue continuum, that at that time  an accretion disk was 
already present, although highly processed. Moreover, the multiple-peaked profiles of 
the emission lines led  to suggest a strong non-homogeneous envelope, 
result of the collision between some outflowing matter and the circumstellar shells.

A few months later, in October, new observation revealed (Yoo 2007) a rapidly developing 
framework, with single-peaked \Ha and \Hb emissions and no absorption 
component at their line centres. The V/R ratio became $<<1$, indicating that the 
accretion disk might have been destroyed within the April-October time interval. 
Variability in the \Ha blue component intensity has been explained as the actual 
expansion effect of some jets. Moreover, the blue continuum became so dim  as to unveil the 
M giant even at short wavelengths, while a contemporary enhancement of the emission lines 
allowed to detect conspicuous [OIII] 4959 and [OIII] 5007, along with broad emission 
components of [NeIII] lines. This emission might have originated in an extended, 
expanding shell constructed by the  outflowing matter from the WD.

In the X-ray, Mukay et al. (2006), analysing  Suzaku observations of 2006 January and May, 
found CH Cyg in a soft X-ray bright, hard X-ray faint state. 
While the soft X-ray flux level was similar to the 1994 ASCA observation, the hard one was 
much lower, confirming that the hard component is more variable than the soft component 
(Ezuka et al. 1997), which in turn may be  spatially distinct. 
Assuming a binary period of $\sim$14.5 years, the entire binary should be 
extended $<$0.4''� (100AU at $\sim$270 pc), explaining why Chandra was not able to resolve 
the system and, at the same time,   in agreement  with an extended X-ray emission region, 
comparable with the size of the binary, as expected in the colliding wind interpretation 
demonstrated  by consistent modelling. 

The hard X-ray component, more directly sensitive to the variable symbiotic activity,  
is the manifestation of those non-thermal emission processes (such as synchrotron) 
which once again prove the key-role played by shocks in accelerating relativistic 
particles responsible for the bulk of emission  at h$\nu>$10 (De Becker et al. 2007).

In Fig. \ref{fig:2007} we present the modelling of the UBV and JHKLM data presented by 
Skopal et al. (2007) and Taranova \& Shenavrin (2007), respectively. 
The models which best fit the slopes of the data are the combination of those used in 
previous epochs (see caption of Fig. \ref{fig:2007}). 
Three main slopes can be noticed. 
That at maximum of  U (red) shows a contribution of two nebulae with different densities. 
The curve with the highest V (green) have a conspicuous contribution from the giant bb radiation. 
This contribution decreases and even disappears at minimum (blue curve), 
indicating that the whole binary system 
is obscured by the expanding nebula, as was found in previous epochs.

\section{Discussion and concluding remarks}
The  complex nature of the SS CH Cygni has been investigated by modelling  the line and 
continuum  spectra at different epochs, from the  impressive outburst of 1978 
up to date. Our results give an approximated picture of the
shocked nebulae throughout the system because our approach  implies
a plane parallel geometry  for the different shock fronts. 
This is constrained by the
composite models which account for both photoionization from the WD and shocks
created by the collision of the winds. However, the plane-parallel
geometry can be justified for relative large radii and in particular situations
characteristic of SS.

The present analysis has revealed that  similar developing trends may be recognised 
throughout the active  phases and throughout the quiescent phases. 
In the light of a colliding-wind scenario, we have specifically focused on the role 
of the nebulae created by collision of the winds from the hot and the cool stars, 
in order to build up a comprehensive view of the system as a whole, namely to self-consistently 
interpret the variable line spectra, the radio emission, the IR photometric properties, 
the fluxes in the U B V bands and the X-ray observations. 
 The most important result of our method  that can be applied to
 all the SSs, consists in
recognizing  the different sources of radiation (stars, nebulae, dust) 
throughout the continuum SED.

We have found that bremsstrahlung from the nebula downstream of the  shock front 
facing the WD between the stars  explains the continuum 
slopes observed during the active states. 
These fast shocks seem to be always present during the outbursts,  proving that 
the wind collision is playing a primary role in the spectral properties of the 
symbiotic phenomenon (e.g. radio and X-ray wavelength region). 
The circumstellar dusty shells define the optic-UV SED , also contributing to 
the line spectra during the active phases.

The WD may be related to flickering during the active phases, either by appearing 
directly (1991-95) or by photoionising the shells throughout a disrupted medium (1978-87). 
Towards the end of the active phases, the expanding shock propagating outward the 
system is visible both in the continuum SED and in the line spectra.

During the quiescent phases  the circumbinary expanding shocks  dominate. 
We have found that these shocks propagate  throughout a relatively dense medium which 
may be interpreted as matter ejected in previous outbursts. 
This leads to a relative strong  synchrotron radiation   from the shock front at some epochs.

The jets observed by Corradi et al (2001)   imply the presence of an 
accretion disk. The modelling of the low-ionisation lines (e.g. [NII, [OII]), 
that are strong in their images, confirms that the jets expanding in the ISM, have 
reached a radius $>$ 0.01 pc from the SS.

We have shown,  by consistent modelling,  that soft and hard X-rays are emitted 
downstream of shock fronts 
corresponding to quite different velocities, namely, the expanding shock 
(\Vs $\sim$ 100-150 \kms) propagating  outwards the binary system would explain 
the soft X-ray emission, while the hard X-rays  correspond to the high velocity 
shock between the stars, in agreement with the recent results discussed 
by Karovska et al. (2007) on the basis of radio, optical and X-ray imaging.

Dust grains within the nebulae are collisionally 
heated by the  gas in a 
strong shock velocity regime ($\geq$ 600 \kms), while, in the dusty shells, grains are 
mainly heated by radiation 
from the hot star. The dust expanding  disrupted shells might reach, in some cases, 
the external edge 
of the symbiotic system (r$>$ 10$^{14}$ cm),  becoming circumbinary and therefore 
contributing to the complex obscuration episodes of CH Cyg.

It should be emphasized that the interpretation of CH Cyg spectral evolution 
is further complicated by the relatively short orbital period ($\sim$14.5 years): 
in fact, the orbital motion combined with the unpredictable recurrent activity leads 
to  different system configuration as seen from Earth. We have tried to explain them 
by  combining the radiation fluxes from the several emitting contributors with different, 
variable weights.

For the sake of clarity, we schematically summarise the results 
of our modelling:
\subsection*{1978-1987 active phase}
\begin{itemize}
\item Emission lines with FWHM of 400 \kms - 1200 \kms correspond to the reverse shock;
\item before 1984 November, low-ionisation lines come from the shells;
\item after 1994, higher level lines are emitted both by the expanding nebula and the dusty shells;
\item at days JD2446212 and JD2446776,  UV and optical lines from the expanding nebulae appear;
\item at the end of the active phase, shock dominated models explain both line and continuum spectra.
\end{itemize}
\subsection*{1988-1990 quiescent phase}
\begin{itemize}
\item The  expanding nebula  propagates throughout  matter ejected in previous eruptions;
\item in the IR the system is obscured independently from wavelength.\\
\end{itemize}
\subsection*{1991-1995 active phase}
\begin{itemize}
\item In 1991, at the onset of the new active state, the lines display FWHM$<$ 100 \kms, 
indicating a disrupted expanding nebula;
\item in August 1991, broad \Ha and \Hb are seen in absorption, suggesting that the reverse 
shock is absorbing the WD flux;
\item by 1992 absorption lines disappear and M-giant molecular bands are observed. 
In May 1993,  UV lines  are emitted from an expanding nebula with a radius of 
10$^{16}$ - 10$^{17}$ cm.
\end{itemize}
\subsection*{1996-1997 quiescent phase}
\begin{itemize}
\item The radio flux shows a synchrotron component created by the Fermi mechanism at 
the shock front.
\end{itemize}
\subsection*{1998-2001 active phase}
\begin{itemize}
\item  . A fast reverse shock of 1200 \kms may explain the X-ray  Chandra observations in 2001;
\item the HST images by Eyres et al. (2002) reveal strong [OIII], which can be 
explained by the expanding nebulae   in agreement with  the  fit of the SED;
\item in 1998 Sept the OI, HeI, HeII, FeII, and TiII lines  are emitted from the shells.  
Moreover, [OIII] 5007 is evident and [OIII] 4363 is  absent both in and out of the eclipse. 
These lines come from    shock-dominated expanding nebulae  with relatively low velocities 
($\sim$70-100 \kms) and relatively low densities (\n0$\sim$ 5000 \cm3).
In fact  the corresponding [OIII] 4363 lines are relatively low;
\item the 1999 images of Corradi et al. (2001) indicate strong [OII] and 
[NII] lines: they are characteristic of  a shock dominated nebula with a 
relative low \n0=5 10$^3$ \cm3. 
\end{itemize}
\subsection*{2001-up to date}
\begin{itemize}
\item three different slopes are observed in the optical-IR range.
They are modelled by  the combination of different expanding models
with the bb flux from the cool component star.
\item the decline observed by Taranova \& Shenavrin in the IR
is consistent with
the decline in the optical-UV range and is explained by the 
eclipse of the whole system.
\end{itemize}
\section*{Acknowledgements}
We are  grateful to Elia Leibowitz  for
interesting conversations. We would like to thank the anonymous referee for many helpful comments.

\begin{appendix}
\section{Observational references}
\begin{table}
\centering \caption{Observation references for the 1978-86 active phase.\label{tab:refa1}}
\small{
\begin{tabular}{lllllllllllllllllll}\\
\hline 
\hline 
& Observation  &  References\\
&date &\\
\hline
radio & 1984 Apr - 1985 May & Taylor et al. (1985, 1986, 1988)\\
IR & 1983 & IRAS CPS$^a$ \\
& 1978-82 &  Ipatov et al. (1984)\\
& 1978-87 & Taranova \& Yudin (1988)\\
optical & 1977-86 & Hack et al. (1988)\\
& 1978-82 &  Ipatov et al. (1984)\\
& 1978-87 &  Taranova \& Yudin (1988)\\
& 1980-87 &  Bondar \& Shakhovskaya (2001)\\
UV &  1978-87  & Taranova \& Yudin (1988)\\
& 1981 Nov 29 &   Mikolajewska et al. (1988)\\
& 1985 Jan 24 &  Mikolajewska et al. (1988)\\
& 1986 Jul 16 &  Mikolajewska et al. (1988)\\
X-rays & 1985 May 24 & Leahy \& Taylor (1987)\\
\hline
\end{tabular}
}
\end{table}

\begin{table}
\centering \caption{Observation references for the 1987-90 quiescent phase. \label{tab:reft1}}
\small{
\begin{tabular}{lllllllllllllllllll}\\
\hline 
\hline 
& Observation  &  References\\
&date &\\
\hline
radio & 1989 Mar 7-12 & Altenhoff et al. (1994)\\
& 1988 Oct 14 & Ivison et al. (1991)\\
IR & 1988-90 &  Taranova \& Yudin (1992)\\
optical & 1988-90 &  Taranova \& Yudin (1992)\\
UV & -  & -\\
X & -  & -\\
\hline
\end{tabular}
}
\end{table}

\begin{table}
\centering \caption{Observation references for the 1991-95 active phase.\label{tab:refa2}}
\small{
\begin{tabular}{lllllllllllllllllll}\\
\hline 
\hline 
& Observation &  References\\
&date&\\
\hline
radio & 1992 Jul 24 &  Skopal et al. (1996)\\
& 1993 Jun 17 - Aug 19  & Skopal et al. (1996)\\
& 1993 &  Skopal et al. (1996)\\
IR & 1991-95 &  Munari et al. (1996)\\
optical & 1989-95 &  Skopal et al. (1996)\\
& 1989-94 & Skopal et al. (1996)\\
&  1991-95 & Munari et al. (1996)\\
UV & 1989-95 & Skopal et al. (1996)\\
& 1991-95 & Munari et al. (1996)\\
& 1992 Nov 24 &  Skopal et al. (1996)\\
& 1993 May 27 &  Skopal et al. (1996)\\
X-rays & 1994 Oct 19 & Ezuka et al. (1997)\\
\hline
\end{tabular}
}
\end{table}

\begin{table}
\centering \caption{Observation references for the 1996-97 quiescent phase.\label{tab:reft2}}
\small{
\begin{tabular}{lllllllllllllllllll}\\
\hline 
\hline 
& Observation &  References\\
&date& \\
\hline
radio & 1997 Mar 8 - Aug 31 &  Karovska et al. (1998)\\
IR & 1997 May 10 & Schild et al. (1999)\\
optical & 1995 Dec 18 - 1997 Mar 11 &  Skopal et al. (1997)\\
& 1996-97 &  Corradi et al. (2001)\\
UV & 1997 Jul 10 & Corradi et al. (2001)\\
X & - & - \\
\hline
\end{tabular}
}
\end{table}

\begin{table}
\centering \caption{Observation references for the 1998-2000 active phase.\label{tab:refa3}}
\small{
\begin{tabular}{lllllllllllllllllll}\\
\hline 
\hline 
& Observation &   References\\
&date& \\
\hline
radio & 1999 Sep 26 &Eyres et al. (2002)\\
IR & 1999-2003 &  Taranova \& Shenavrin (2004)\\
optical & 1998 Sep 5 - 2000 Oct 15 &  Eyres et al. (2002) \\
& 1998 Jun 29 - 2001 Nov 28 & Skopal et al. (2000, 2002)\\
& 1999 Aug 12 &  Eyres et al. (2002)\\
UV & 1999 Aug 12 &  Eyres et al. (2002)\\
& 1998-2000 Apr &  Eyres et al. (2002)\\
X-rays & - & - \\
\hline
\end{tabular}
}
\end{table}

\begin{table}
\centering \caption{Observation references for the most recent period 2001-07.\label{tab:reft3}}
\small{
\begin{tabular}{lllllllllllllllllll}\\
\hline 
\hline 
& Observation &  References\\
&date& \\
\hline
radio & - & - \\
IR & 1999-2007 &  Taranova \& Shenavrin (2007)\\
optical & 2001 Nov 28 - 2003 Nov 2003 &  Skopal et al. (2004)\\
&2003 Nov 8 - 2007 Jan 15  & Skopal et al. 2007\\
&2004 Apr 9-11  & Yoo (2006)\\
&2004 Oct 2-5  & Yoo (2007)\\
UV &  - & - \\
X & 2001 Mar 27 &  Galloway \& Sokoloski (2004)\\
& 2006 Jan 04/05 - May 28/29 & Mukai et al. (2006)\\
\hline
\end{tabular}
}
\end{table}
\end{appendix}
\end{document}